\documentstyle[twocolumn,epsf]{mn} 
\title[Self-consistent axisymmetric Sridhar--Touma models] 
      {Self-consistent axisymmetric Sridhar--Touma models}
\author[M. A. Jalali and P. T. de Zeeuw] 
	{Mir Abbas Jalali$^1$\thanks{jalali@iasbs.ac.ir}  
	and P.\ Tim de Zeeuw$^2$\thanks{tim@strw.leidenuniv.nl} \\ 
	$^1$Institute for Advanced Studies in Basic Sciences,  
	    P.O.\ Box 45195-159, Zanjan, IRAN \\ 
	$^2$Sterrewacht Leiden, Postbus 9513, 2300 RA Leiden, 
	    The Netherlands} 
 
\begin{document} 
\label{firstpage} 
\maketitle 
 
\begin{abstract}  
We construct phase-space distribution functions for the oblate, cuspy
mass models of Sridhar \& Touma, which may contain a central point
mass (black hole) and have potentials of St\"ackel form in parabolic
coordinates. The density in the ST models is proportional to a power
$r^{-\gamma}$ of the radius, with $0<\gamma<1$. We derive distribution
functions $f(E, L_z)$ for the scale-free ST models (no black hole)
using a power series of the energy $E$ and the component $L_z$ of the
angular momentum parallel to the symmetry axis. We use the contour
integral method of Hunter \& Qian to construct $f(E, L_z)$ for ST
models with central black holes, and employ the scheme introduced by
Dejonghe \& de Zeeuw to derive more general distribution functions
which depend on $E$, $L_z$ and the exact third integral $I_3$. We find
that self-consistent two- and three-integral distribution functions
exist for all values $0 < \gamma < 1$.
\end{abstract} 
 
 
\section{Introduction}
\label{sec:introduction}
 
Observations of the nuclei of elliptical galaxies using the Hubble
Space Telescope have revealed that the luminosity density diverges
towards the centre as a power $r^{-\gamma}$ of the radius
\cite{J94,L95,C97,R01}. Many and perhaps all of the nuclei host a
central black hole \cite{R99,Z01}. Dynamical models for such cusped
galaxies include the axisymmetric power-law galaxies introduced by
Toomre (1982) and Evans (1994), which have spheroidal potentials and
simple $f(E, L_z)$ distribution functions, with $E$ the orbital energy
and $L_z$ the component of the angular momentum parallel to the
symmetry axis. Similar models with spheroidal densities were
constructed by Dehnen \& Gerhard (1994) and Qian et al.\ (1995,
hereafter Q95). Most orbits in these models admit an approximate third
integral of motion. Self-consistent three-integral distribution
functions were constructed for some special scale-free cases by
Schwarzschild's (1979) numerical orbit superposition method (Richstone
1980, 1984; Levison \& Richstone 1985) and by analytic means (de
Zeeuw, Evans \& Schwarzschild 1996; Evans, H\"afner \& de Zeeuw 1997,
hereafter EHZ).\looseness=-2

Q95 constructed two-integral distribution functions $f(E, L_z)$ for
axisymmetric power-law spheroidal densities containing a central black
hole by means of the contour integral method of Hunter \& Qian (1993,
hereafter HQ). They showed that self-consistent $f(E, L_z)$'s exist
for oblate scale-free spheroids with $0 \leq \gamma < 3$, but found an
upper bound for the axial ratios of self-consistent prolate models of
this kind. Their study also revealed that the presence of a central
black hole limits the region of parameter space for which physical
distribution functions (i.e., $f(E, L_z) \geq 0$) exist to $\gamma
\geq 1/2$ (for all axis ratios). De Bruijne et al.\ (1996) studied
oblate spheroidal cusps in the radial range where the spherical
potential of the black hole dominates, and constructed analytic
constant-anisotropy three-integral distribution functions $f(E, L_z,
L)$, with $L$ the modulus of the angular momentum. These distribution
functions remain physical also when $0 \leq \gamma < 1/2$, and suggest
that shallow-cusped axisymmetric galaxies with central black holes can
in fact be constructed, a result confirmed by application of
Schwarzschild's numerical method (Verolme, priv.\ comm.).
Self-consistent axisymmetric models with central black holes designed
to fit the observed photometry and kinematics of specific galaxies
were constructed with Schwarzschild's method by, e.g., van der Marel
et al.\ (1998), Cretton et al.\ (1999, 2000) and Gebhardt et al.\
(2000).

The oblate models introduced by Sridhar \& Touma (1997a, hereafter ST)
provide the only set of cuspy models with a central black hole for
which all orbits have an exact third integral of motion $I_3$, so
that, in principle, exact distribution functions can be constructed.
The potential of these models is of St\"ackel form in parabolic
coordinates, and this causes the density to be significantly
flattened, with a shape that is fixed once the cusp slope is chosen,
and a total mass that is infinite. The oblate axisymmetric models with
a potential of St\"ackel form in prolate spheroidal coordinates
\cite{K56,Z85} also admit an exact third integral and include a large
set of models with a range of density profiles and shapes with finite
total mass. However, they all have constant density cores and can not
contain a central point mass without destroying the separability
\cite{Z86}.  Oblate density distributions with potentials that are
separable in spherical coordinates have densities that become negative
at large distances (Lynden--Bell 1962b). Here we study self-consistent
distribution functions for the oblate ST models. We show that, by
contrast to the cusped spheroidal densities of Q95, the ST models with
a central black hole do have consistent $f(E, L_z)$ distribution
functions for all values of the cusp slope $0 < \gamma < 1$.

In \S\ref{sec:st-models} we summarize the properties of the ST models.
Without a black hole, the ST models are scale-free. We discuss their
distribution functions in \S\ref{sec:no-bh}, and investigate the
general case in \S\ref{sec:with-bh}.  We summarize our conclusions in
\S\ref{sec:conclusions}.\looseness=-2

\section{Sridhar--Touma models} 
\label{sec:st-models}

A comprehensive description of the mass models and the orbit structure
can be found in ST. Here we collect the relevant properties, derive
the fundamental integral equation for the distribution function, and
close with a brief discussion of the velocity moments.

\subsection{Potential, density and orbits}
\label{sec:st-models-basics}
 
The motion in the axisymmetric ST models separates in parabolic
coordinates $(\xi, \eta)$ in the meridional plane.  We define them as
\begin{equation} 
\label{eq:parabolic-coordinates} 
\xi = r(1+\cos \theta), \quad 
\eta = r(1-\cos \theta), \quad 
\xi, \eta \ge 0,  
\end{equation} 
where $r$ is the polar radius and $\theta$ is the co-latitude. Our
definition of $\eta$ differs from that in ST by an overall sign, which
removes the need to use $|\eta|$ in many expressions. In these
coordinates the gravitational potential of an ST model can be written
as
\begin{equation} 
\label{eq:potential-parabolic} 
V(\xi,\eta) = \frac {2K(\xi ^{3-\gamma}+\eta ^{3-\gamma}) -2GM} {\xi +\eta},  
	      \quad K>0,  
\end{equation} 
where $K$ is a positive constant, $G$ is the gravitational constant,
and $M$ denotes the mass of a central black hole (point mass). The
density distribution $\rho(\xi,\eta)$ associated with the potential
(\ref{eq:potential-parabolic}) follows from eq.\ (18) of ST.  When
$M=0$, the potential and the density are proportional to
$r^{2-\gamma}$ and $r^{-\gamma}$, respectively, i.e., they are
scale-free, but the density is non-negative only for $0<\gamma <1$.
The equipotential surfaces are approximately spheroidal. Their axis
ratio $b/a$ (defined by the condition $V(0,b)=V(a,0)$) equals 1/2 for
all values of $\gamma$. As a result, the surfaces of constant density
are dimpled along the short ($z$) axis, and the dimple deepens, i.e.,
the density distribution becomes increasingly toroidal, with
increasing $\gamma$.

For subsequent use, we record here the expressions for the ST
potential-density pairs in terms of the standard spherical polar
coordinates $(r, \theta, \phi)$ 
\begin{eqnarray} 
\label{eq:pot-spherical}
V(r,\theta) \!\!\!&=& \!\!\! Kr^{2-\gamma}P(\theta)-\frac{GM}{r}, \nonumber \\
\rho (r,\theta) \!\!\! &=& \!\!\! \rho _0 r^{-\gamma}S(\theta) +M\delta(r),  
\end{eqnarray} 
where $\rho_0=2(3-\gamma)(1-\gamma)K/\pi G$, $\delta$ is
the Dirac delta-function, and 
\begin{eqnarray}
\label{eq:dens-spherical} 
P(\theta) \!\!\!&=& \!\!\!(1\!+\!\cos\theta)^{3-\gamma}+ 
		    (1\!-\!\cos\theta)^{3-\gamma}, \nonumber \\  
S(\theta) \!\!\!&=& \!\!\!(2\!-\!\gamma \!-\!\cos \theta) 
		    (1\!+\!\cos\theta)^{2-\gamma} \nonumber \\
		&\null& \!\!\!+(2\!-\!\gamma \!+\!\cos\theta)  
		    (1\!-\!\cos\theta)^{2-\gamma}. 
\end{eqnarray} 
In terms of cylindrical polar coordinates ($\varpi, \phi, z$):  
\begin{eqnarray} 
\label{eq:dens-pot-cylindrical}
V(\varpi, z) \!\!\!\!&=&\!\!\!\! 
		  \frac {K}{r} \big[(r\!+\!z)^{3-\gamma} \!+\! 
				      (r\!-\!z)^{3-\gamma} \big]
		  -\frac{GM}{r},                      \nonumber \\
\rho(\varpi, z) \!\!\!\!&=& \!\!\!\!\frac {\rho_0}{r^3} \big \{ 
	[(2-\gamma) r\!\!-\!\!z] (r\!+\!z)^{2-\gamma} \nonumber \\
	&\null& \!+ 
	[(2-\gamma) r\!+\!z](r\!\!-\!\!z)^{2-\gamma}\big \} \!+\!M\delta(r), 
\end{eqnarray} 
where $r=\sqrt {\varpi^2+z^2}$. In what follows, we take units such
that $K=1$ and $\rho _0=1$.

The separation of the Hamilton-Jacobi equation results in a third
integral of motion given by \cite{P65}
\begin{equation} 
\label{eq:ithree-parabolic} 
I_3=2\xi p_\xi^2+2\xi ^{3-\gamma}+\frac {L_z^2}{2\xi}-E\xi -GM, 
\end{equation} 
or, equivalently  
\begin{equation} 
\label{eq:ithree-eta-parabolic}
I_3=-2\eta p_\eta^2-2\eta ^{3-\gamma}-\frac {L_z^2}{2\eta}+E\eta+GM,  
\end{equation} 
with $p_\xi$ and $p_\eta$ the momenta conjugate to $\xi$ and $\eta$,
respectively. Here $E$ denotes the orbital energy and $L_z\equiv
p_{\phi}=r^2 \dot \phi \, \sin ^2 \theta$ the component of the angular
momentum parallel to the symmetry axis, both of which are integrals of
motion as well.  All orbits with $L_z\not=0$ are short-axis tubes,
which may be either symmetric or asymmetric with respect to the
equatorial plane. Those with $L_z=0$ are confined to a plane of
constant $\phi$, and are centrophilic when a black hole is present.
The orbital families are illustrated in Figure 2 of ST.\looseness=-2

\subsection{Distribution functions} 
\label{sec:st-models-dfs}
 
In integrable systems, the distribution function depends on the
phase-space coordinates through the isolating integrals of motion
(Jeans 1915; Lynden--Bell 1962a), which for axisymmetric systems means
$f=f(E, L_z, I_3)$. The mass density $\rho$ is related to $f$ through
the integral
\begin{eqnarray} 
\label{eq:fund-int-eq}
&\null&\!\!\!\!\!\!\! \rho(r,\theta)= \int \!\!  
		       f(E,\! L_z,\! I_3)\, {\rm d}^3 \bmath{v} \nonumber \\  
&\null& \!\!\! = 8\!\int \!\!\! {f(E,\! L_z,\! I_3) \over r^2\sin \theta}  
      \left| \frac{\partial (\xi,\eta,\phi)} {\partial (r,\theta,\phi)} \!
	     \frac {\partial (p_{\xi},p_{\eta},p_{\phi})} 
      {\partial (E,I_3,L_z)} \right| {\rm d}E{\rm d}L_z{\rm d}I_3, 
\end{eqnarray}
where $\bmath v=(v_r,v_{\theta},v_{\phi})^{\rm T}$ is the velocity
vector in the spherical coordinates, and the factor of eight occurs
because both $E$ and $I_3$ are quadratic in the velocities, and we
consider only distribution functions that are even in $L_z$, so that
$f(E, -L_z, I_3) = f(E, L_z, I_3)$. The Jacobians in the integral
(\ref{eq:fund-int-eq}) follow from expressions
(\ref{eq:parabolic-coordinates}), (\ref{eq:ithree-parabolic}) and
(\ref{eq:ithree-eta-parabolic}), and are given by
\begin{eqnarray}
\label{eq:jacobians} 
\frac{\partial (\xi,\eta,\phi)}{\partial (r,\theta,\phi)} 
			     \!\!\! &=&\!\!\! 2r \sin \theta, \nonumber \\ 
\left| \frac {\partial (p_{\xi},p_{\eta},p_{\phi})} 
	     {\partial (E,I_3,L_z)} \right | \!\!\! &=& \!\!\! 
	\left|-\frac {\xi+\eta} {16 \xi \eta p_{\xi} p_{\eta}}\right| 
							      \nonumber\\
 &=& \!\!\! \frac 1{4 \sin \theta \sqrt{[I_3\!-\!I_3^-] 
					[I_3^+\!-\!I_3]}},  
\end{eqnarray} 
where  
\begin{eqnarray} 
I_3^-=I_3^-(E,L_z,\xi) \!\!\!&=& \!\!\! 
		2\xi^{3-\gamma}+\frac {L_z^2}{2\xi}-E\xi-GM, \nonumber \\ 
I_3^+=I_3^+(E,L_z,\eta) \!\!\!&=&\!\!\! 
		 -2\eta ^{3-\gamma}-\frac {L_z^2}{2\eta}+ E\eta +GM 
\end{eqnarray} 
We define the {\it equivalent surface density} $\Sigma (r,\theta)$ by
\begin{equation}
\label{eq:equiv-surf-den}
\Sigma (r,\theta) =2r\sin \theta \rho(r,\theta),
\end{equation}
and substitute expression (\ref{eq:jacobians}) in the integral
(\ref{eq:fund-int-eq}) to obtain
\begin{equation} 
\label{eq:surf-fund-int} 
\Sigma(r,\theta) = \! 
		 8 \!\!\!\!\int\limits_{V(r,\theta)}^{\infty} 
				 \! \!\!\!\!{\rm d}E \!\!\!\!\!\!
		 \int\limits_{0}^{{L_z}_{\rm max}(E,r,\theta)} 
					   \!\!\!\!\!\!\!\!\!\!\!{\rm d}L_z 
		 \int\limits_{I_3^-(E,L_z,\xi)}^{I_3^+(E,L_z,\eta)}
				       \!\!\!\!\!\!\!\!\!\!{\rm d}I_3 \! 
		 \frac{f(E,L_z,I_3)}{\sqrt{[I_3\!-\!I_3^-] 
					      [I_3^+\!-\!I_3]}}, 
\end{equation}
where
\begin{equation} 
\label{eq:lzmax}
{L_z}_{\rm max}(E,r,\theta)=r \sin \theta  \sqrt {2[E-V(r,\theta)]}.  
\end{equation} 
This is the fundamental integral equation for the distribution
functions of the ST models. According to (\ref{eq:lzmax}), the
distribution functions obtained from (\ref{eq:surf-fund-int}) will be
even in $L_z$. We assign only positive values to ${L_z}_{\rm max}$,
which means that all stellar orbits have a definite sense of rotation
around the axis of symmetry.

\begin{figure*} 
\centerline{\hbox{\epsfxsize=2.1in\epsfbox{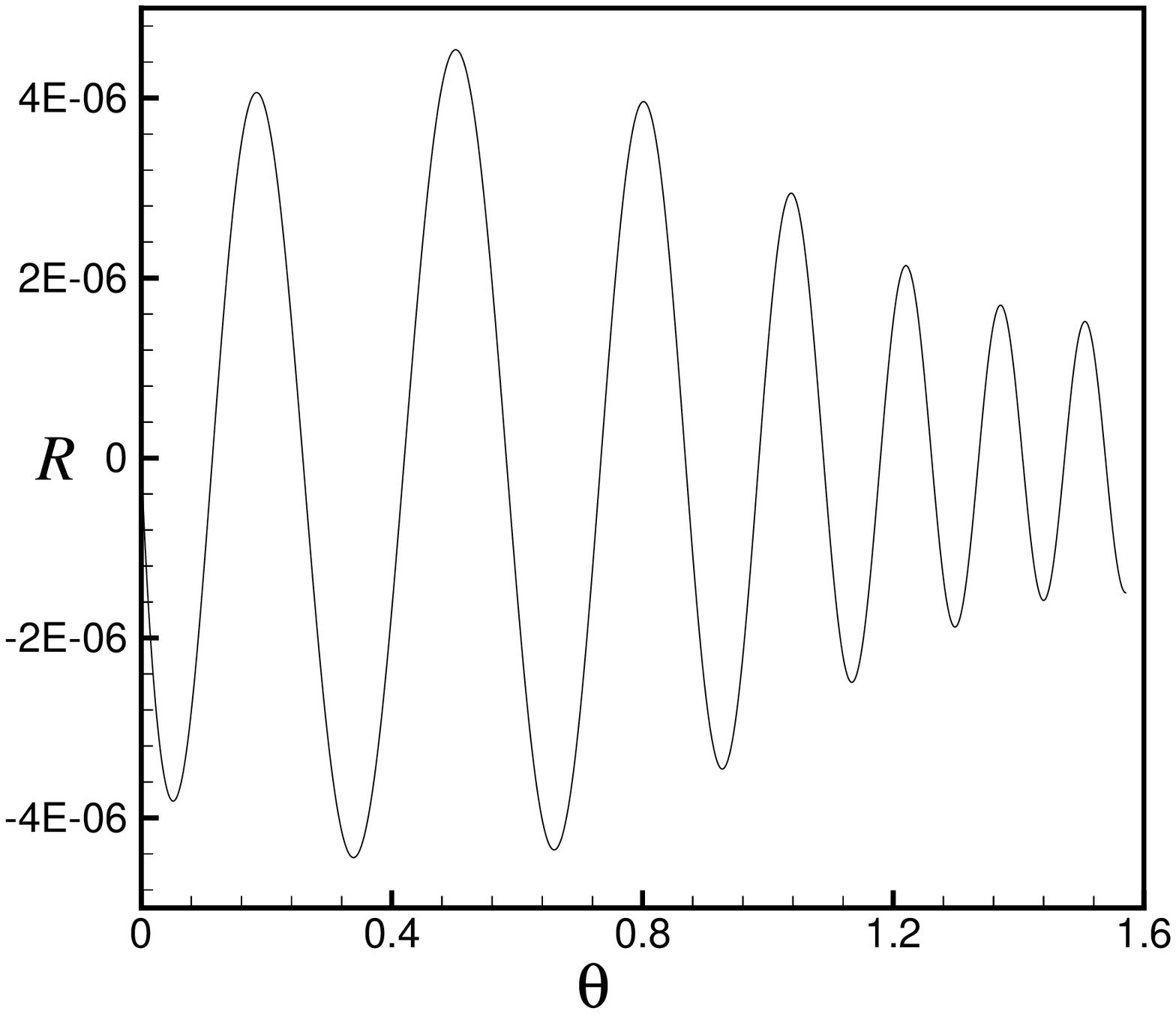}\hspace{0.3cm} 
		  \epsfxsize=2.1in\epsfbox{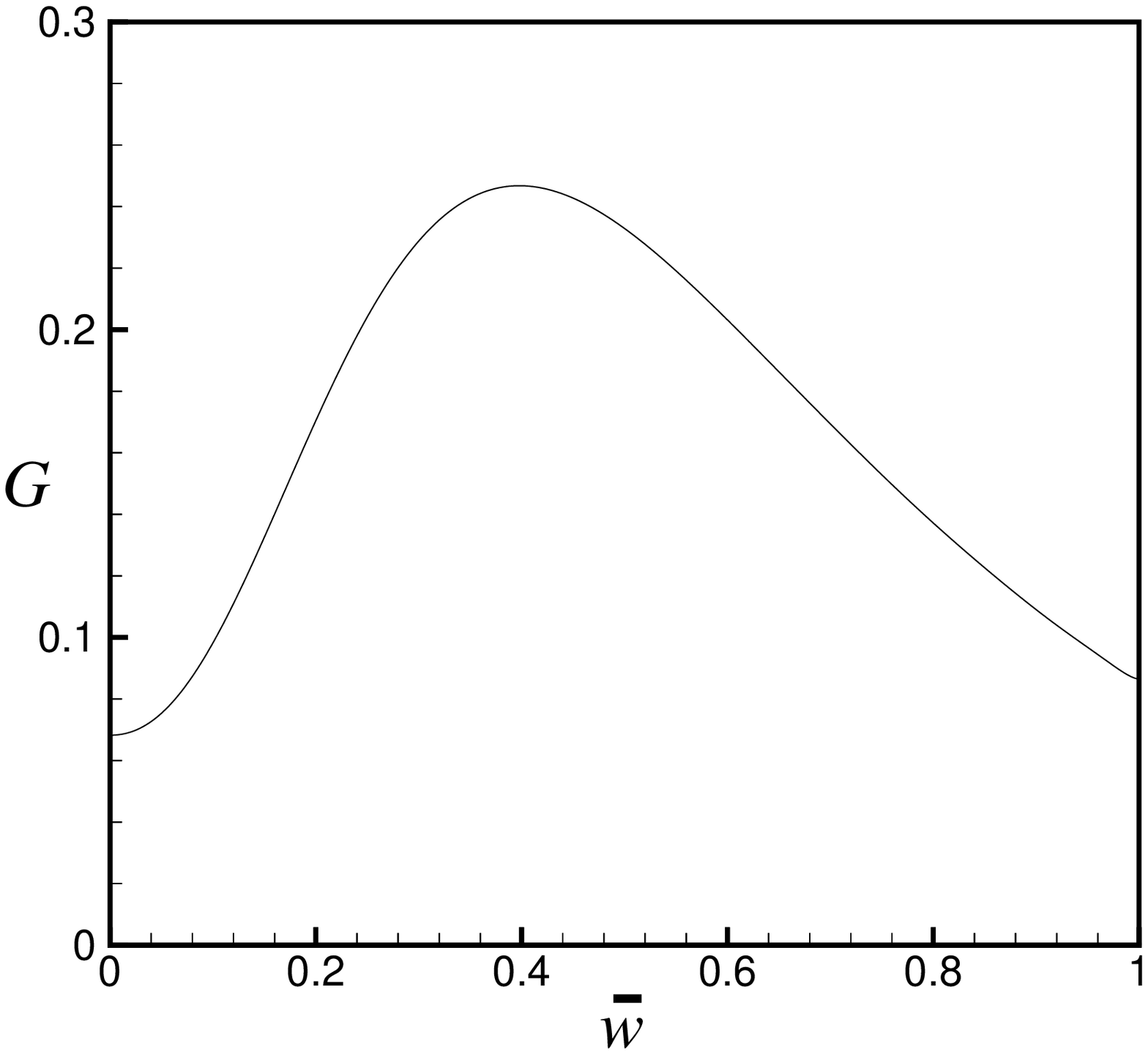}\hspace{0.3cm} 
		  \epsfxsize=2.1in\epsfbox{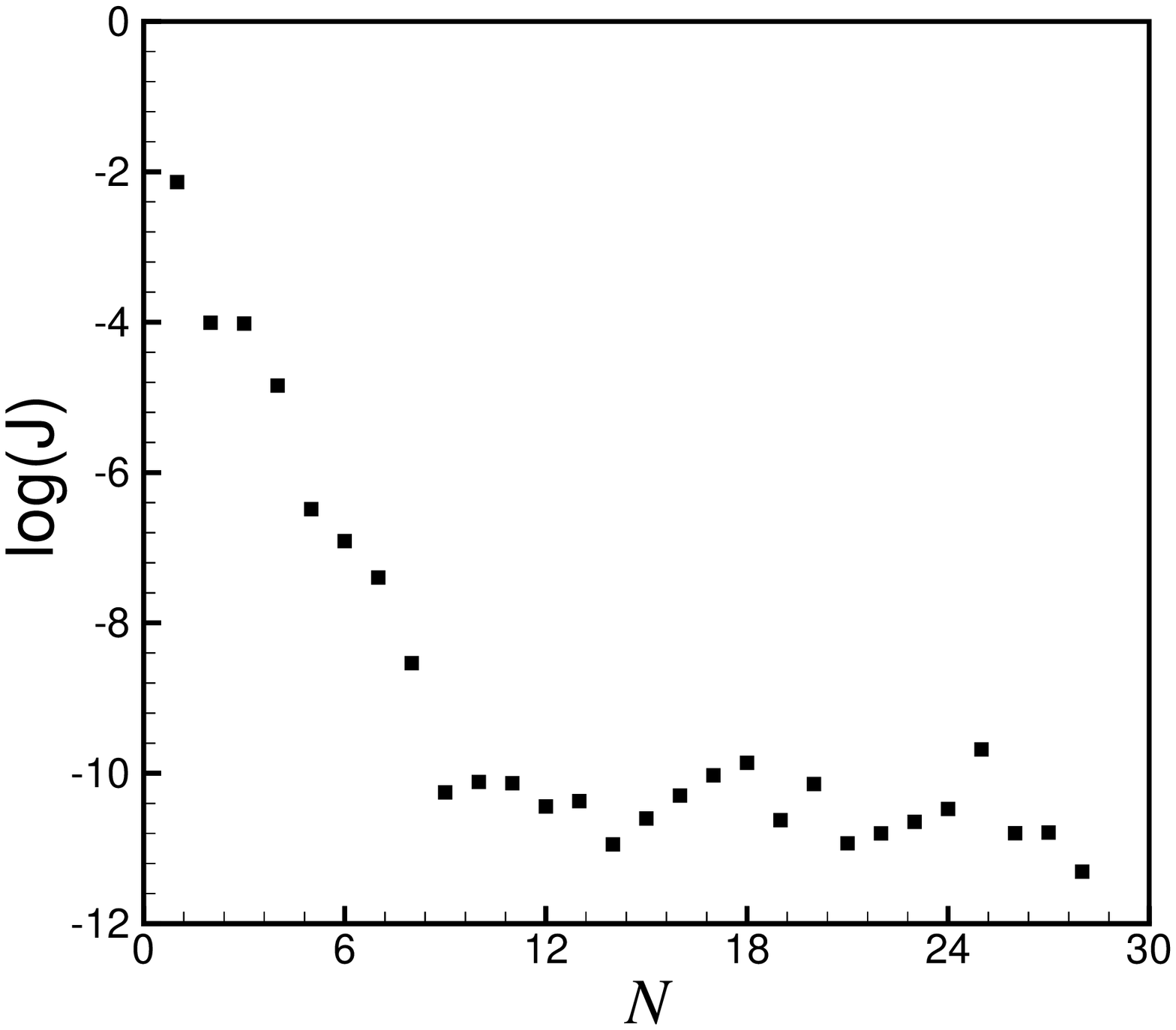}}} 
\centerline{\hspace*{3cm}$(a)$\hfill$(b)$\hfill$(c)$\hspace{1.0in}} 
\caption[Fig1]{(a) The variation of $R(\theta)=\sigma(\theta) -\sigma
_N(\theta)$ for $\gamma=0.7$ and $N=28$.  (b) The distribution
function ${\cal G}(\overline w)$ for $0\le \overline w=w/w_{\rm
max}\le 1$. (c) The history of the objective function ${\cal J}$
versus $N$. ${\cal J}$ has converged to its minimum at $\approx
10^{-12}$. The numerical accuracy is saturated because of truncation
and round-off errors.}
\end{figure*} 

\subsection{Velocity moments}
\label{sec:st-models-moments}

The velocity moments $\rho \langle v_\xi^l v_\eta^m v_\phi^n \rangle$
are defined by
\begin{equation} 
\label{eq:vel-moments-xi-eta}
\rho \langle v_\xi^l v_\eta^m v_\phi^n \rangle \!= \!\!
\int \!\!\int \!\!\int v_\xi^l v_\eta^m v_\phi^n 
		       f(\xi, \eta, v_\xi, v_\eta,v_\phi) \, 
		{\rm d}^3 \bmath{v}, 
\end{equation} 
where the integrations are carried out over all possible velocities.
The velocity components are related to the momenta in parabolic 
coordinates through 
\begin{equation}
\label{eq:parabolic-velocities}
p_\xi  \!=\!{\cal L}^2 \dot \xi  \!=\! {\cal L} v_\xi,~~
p_\eta \!=\!{\cal M}^2 \dot \eta \!=\! {\cal M} v_\eta,~~
p_\phi \!=\!{\cal N}^2 \dot \phi \!=\! {\cal N} v_\phi,
\end{equation}
where ${\cal L}$, ${\cal M}$ and ${\cal N}$ are the metric coefficients 
defined as
\begin{equation}
\label{eq:metric-coefficients}
{\cal L}^2={\xi +\eta \over 4 \xi},~~{\cal M}^2={\xi+\eta \over 4\eta},
~~{\cal N}^2=\xi \eta. 
\end{equation}
Transforming to the integration variables $E$, $L_z$ and $I_3$ gives
(cf.\ eq.\ [\ref{eq:surf-fund-int}])
\begin{eqnarray}
\label{eq:vel-moments-integrals}
&\null& \!\!\!\!\! \rho \langle v_\xi^l v_\eta^m v_\phi^n \rangle =  
2^{2+\frac {l+m}{2}}\xi^{l-n-1 \over 2}\eta^{m-n-1 \over 2}
(\xi+\eta)^{-(l+m)\over 2} \nonumber \\
&\null& \!\!\!\! \times \int \!  
{f(E,L_z,I_3)L_z^n\over (I_3\!-\!I_3^-)^{{1-l} \over 2}
(I_3^+\!-\!I_3)^{{1-m}\over 2}} \, {\rm d}I_3{\rm d}L_z{\rm d}E,
\end{eqnarray}
where the integration limits are identical to those of the integral
(\ref{eq:surf-fund-int}).

We consider only the second moments.  It is not difficult to show that
the velocity ellipsoid is aligned with the parabolic coordinate
system, so that $\langle v_\xi v_\eta \rangle = \langle v_\xi v_\phi
\rangle =\langle v_\eta v_\phi \rangle =0$. This leaves three non-zero
elements of the stress tensor, $\langle v_\xi^2 \rangle$, $\langle
v_\eta^2 \rangle$, and $\langle v_\phi^2 \rangle$, which are connected
to the potential $V$ and the density $\rho$ by two Jeans equations, 
\begin{eqnarray}
\label{eq:jeans-xi-eta}
{\partial \rho \langle v_\xi^2 \rangle \over \partial \xi} \!\!\!\!\!&=&\!\!\!\!\! 
{\rho \over 2\xi (\xi\! +\!\eta)} \big [ \xi \langle v_\eta^2 \rangle
\!+\! (\xi + \eta) \langle v_\phi^2 \rangle  \nonumber \\ 
&\null& - ( 2 \xi + \eta ) \langle v_\xi^2 \rangle \big ] \! - \!
\rho {\partial V \over \partial \xi}, 
							\nonumber \\
{\partial \rho \langle v_\eta^2 \rangle \over \partial \eta} 
\!\!\!\!\!&=&\!\!\!\!\!
{\rho \over 2 \eta (\xi\! +\!\eta)} \big [ \eta \langle v_\xi^2 \rangle
\!+\! (\xi +\eta) \langle v_\phi^2 \rangle  \nonumber \\
&\null& - ( 2 \eta + \xi ) \langle v_\eta^2 \rangle \big ]\!\!-\!\!
\rho{\partial V \over \partial \eta}.
\end{eqnarray}
The relations with the familiar second moments in spherical
coordinates are
\begin{eqnarray}
\label{eq:parabolic-spherical}
\rho\langle v_r^2 \rangle \!\!\!\!&=&\!\!\!\! 
\frac 1{\xi+\eta}\left [\xi \rho\langle 
v_\xi^2 \rangle+\eta \rho\langle v_\eta^2 \rangle \right ], \nonumber \\
\rho\langle v_\theta^2 \rangle \!\!\!\!&=&\!\!\!\! 
\frac 1{\xi+\eta} \left [ \eta \rho\langle v_\xi^2 \rangle +
\xi \rho\langle v_\eta^2 \rangle \right ], \nonumber \\
\rho\langle v_r v_\theta \rangle \!\!\!\!&=&\!\!\!\! 
\frac {\sqrt{\xi\eta}}{\xi+\eta} \left [ \rho\langle v_\eta^2 \rangle -
\rho\langle v_\xi^2 \rangle \right ].
\end{eqnarray} 
These relations are valid for any oblate density $\rho$ in a
potential $V$ that is separable in parabolic coordinates.
In the two-integral limit, we have $\langle v_r^2 \rangle=\langle
v_\theta^2 \rangle$ and $\langle v_r v_\theta \rangle=0$, or,
equivalently, $\langle v_\eta^2 \rangle = \langle v_\xi^2 \rangle$. 

Using (\ref{eq:vel-moments-integrals}) one can show that the second
velocity moments for the self-consistent ST models are all
infinite. The problem occurs at $E=\infty$. For the case $M=0$ it also
follows through the application of eq.\ (2.14) of EHZ, which gives a
divergent integral for the $\theta$-dependence of the stresses, and
confirms that this property is shared by all weak cusps with $0<\gamma
<1$. Scale-free steep cusps with $\gamma \ge 1$ do not have this
problem, because their density falls off sufficiently fast with
radius.

\section{Scale-free ST models}
\label{sec:no-bh}

In the absence of a central black hole, the ST models are scale-free,
and it is natural to consider distribution functions which are also
scale-free. This restriction simplifies the fundamental integral
equation, as we show in this section.

\subsection{Fundamental integral equation}
\label{sec:no-bh-fie}

When $M=0$, the integrals $L_z$ and $I_3$ can be scaled by the energy
$E$. We consider scale-free distribution functions $f(E, L_z, I_3)$ of
the form
\begin{equation} 
\label{eq:DF-ansatz}
f(E,L_z,I_3)=E^p{\cal G}(E^q I_3, E^s L_z),  
\end{equation} 
where the exponents $p$, $q$ and $s$ are real numbers.  We define
the dimensionless variables
\begin{equation} 
\label{eq:scaled-integrals}
u=\frac{V}{E}, \quad v=I_3E^q,   \quad w=L_zE^s, 
\end{equation}  
from which we obtain
\begin{equation} 
{\rm d}E=-E\frac {{\rm d}u}u, \quad 
{\rm d}I_3=E^{-q} {\rm d}v,   \quad
{\rm d}L_z=E^{-s}{\rm d}w. 
\end{equation} 
We choose the values of $p$, $q$ and $s$ so that the radius $r$
factors out of the integral equation (\ref{eq:surf-fund-int}). This
occurs when:
\begin{equation}
\label{eq:pqs}
p=-\frac 12-\frac 2{2-\gamma},~~
q=-1-\frac 1{2-\gamma},~~
s=-\frac 1{2-\gamma}-\frac 12. 
\end{equation}
It follows that scale-free distributions for the ST models are of the
form (\ref{eq:DF-ansatz}), with $p$, $q$ and $s$ given in
(\ref{eq:pqs}). 

In terms of the new variables (\ref{eq:scaled-integrals}), equation
(\ref{eq:surf-fund-int}) becomes
\begin{equation} 
\label{eq:fund-int-scaled}
\sigma(\theta)\!=\!\!\int\limits_{0}^{1} \!\! u^{2\gamma-3 \over 2-\gamma}\, 
		 {\rm d}u \!\!\!\!\int\limits_{0}^{w_+(u,\theta)} 
				  \!\!\!\!\!\!{\rm d}w  
		 \!\!\int\limits_{v_-(u,w,\theta)}^{v_+(u,w,\theta)}
				   \!\!\!\!\!\!\!\!\!\!{\rm d}v  \, 
\frac{{\cal G}(v,w)}{\sqrt {[v\!-\!v_-][v_+\!-\!v]}}, 
\end{equation} 
where the integration limits are given by   
\begin{eqnarray}
\label{eq:vw-int-limits} 
w_+(u,\theta) \!\!\!\!&=& \!\!\!\!\sqrt {2(1\!-\!u)} u^{\frac 1{2-\gamma}} 
			 \frac{\sin\theta}{P(\theta)^{\frac 1{2-\gamma}}}, 
							      \nonumber \\ 
v_+(u,w,\theta) \!\!\!\!&=&\!\!\!\! 
			    \frac{(1\!-\!\cos\theta)u^{\frac 1{2-\gamma}}} 
			    {P(\theta)^{\frac {3-\gamma}{2-\gamma}}} 
			    \Big[P(\theta)\!-\! 
				2u(1\!-\!\cos \theta)^{2-\gamma}  \nonumber\\
 &\phantom{=}& \qquad\qquad - w^2u^{-\frac 2{2-\gamma}} 
			     \frac{P(\theta)^{4-\gamma \over 2-\gamma}} 
			      {2(1\!-\!\cos \theta)^2} \Big], \nonumber \\
v_-(u,w,\theta)\!\!\!\! &=& \!\!\!\! 
			    \frac{(1\!+\!\cos \theta)u^{\frac 1{2-\gamma}}} 
			     {P(\theta)^{\frac {3-\gamma}{2-\gamma}}} 
			   \Big[ 2u(1\!+\!\cos \theta)^{2-\gamma} \!-\! 
				  P(\theta)\! \nonumber\\
 &\phantom{=}& \qquad\qquad + w^2u^{-\frac 2{2-\gamma}} 
				  \frac{P(\theta)^{4-\gamma \over 2-\gamma}} 
				  {2(1\!+\!\cos \theta)^2} \Big], 
\end{eqnarray} 
and the left hand side is defined as
\begin{equation} 
\sigma(\theta) = \frac {2S(\theta)\sin \theta} 
{P(\theta)^{1-\gamma \over 2-\gamma}}.  
\end{equation} 
Hence, the problem of finding $f(E,L_z,I_3)$ is reduced to solving
equation (\ref{eq:fund-int-scaled}) for ${\cal G}(v,w)$.

\subsection{Fricke series for $f(E, L_z)$} 
\label{sec:no-bh-two}

We first consider the (even part of the) two-integral distribution
function $f(E, L_z)$, which is determined uniquely by the density
distribution. One way to obtain it is to assume that ${\cal G}$ only
depends on $w$. In this case equation (\ref{eq:fund-int-scaled}) reads
\begin{equation} 
\label{eq:fund-twoint-scaled}
\sigma(\theta)=\pi \int\limits_{0}^{1} \!\! u^{2\gamma -3 \over 2-\gamma} 
	       {\rm d}u \!\!\! \int\limits_{0}^{w_+(u,\theta)} \!\!\!\! 
		    {\cal G}(w){\rm d}w.   
\end{equation} 
This integral equation must be solved for ${\cal G}(w)\ge0$.  We
assume the solution in the form of power series (Fricke 1952)
\begin{equation} 
\label{eq:fricke-series-g}
{\cal G}(w)=\sum _{n=0}^{N}a_n w^n, 
\end{equation} 
where we have to determine $a_n$ ($n=0,\ldots, N$) and we choose $N$
based on the required accuracy of the solutions.  We substitute the
series (\ref{eq:fricke-series-g}) into the integral equation
(\ref{eq:fund-twoint-scaled}) and obtain
\begin{equation} 
\label{eq:fricke-series-s}
\sigma_N (\theta) = \sum _{n=0}^{N} a_n g_n(\theta),  
\end{equation} 
where the functions $g_n(\theta)$ are given by (Gradshteyn \& Ryzhik
1980)
\begin{eqnarray} 
g_n(\theta) \!\!\!\! &=& \!\!\!\! 
		       \pi \int\limits_{0}^{1} \!\! 
			    u^{2\gamma-3 \over 2-\gamma}  {\rm d}u \!\!\!
		       \int\limits _{0}^{w_+(u,\theta)} \!\!\!\!w^n {\rm d}w 
					   \nonumber \\
		   &=& \!\!\!\!  \frac{\pi}{n\!+\!1} 
		      \left ( \frac {\sqrt{2} \sin \theta} 
		      {P(\theta)^{1\over 2-\gamma}} \right )^{n+1} 
		      {\rm B}\left (\frac{n\!+\gamma}{2-\gamma},
					\frac{n\!+\!3}{2} \right ),
\end{eqnarray} 
and $B$ is the Beta-function.  We determine the $a_n$ from the
requirement that $\sigma_N(\theta)$ converges to $\sigma(\theta)$ as
$N$ is increased. We think of mean convergence and therefore attempt to
minimize the objective function
\begin{equation} 
{\cal J}=\int\limits _{0}^{\frac {\pi}2} 
{\textstyle {1\over 2}}  \left [\sigma(\theta)- \sigma_N(\theta) 
\right ]^2 {\rm d} \theta,  
\end{equation} 
with respect to the variations of the coefficients $a_n$. This is the
well-known method of Bubnov--Galerkin (e.g., Reddy 1986). The function
${\cal J}$ has a local extremum if
\begin{equation} 
\frac {\partial {\cal J}}{\partial a_j}=0, \qquad j=0,\ldots, N. 
\end{equation} 
Therefore, we are left with a set of linear algebraic equations  
for $a_n$ as 
\begin{equation} 
\bmath {K} \cdot \bmath {a}=\bmath {b},  
\end{equation} 
where the matrix $\bmath {K}=[k_{ij}]$ is the so-called {\it
stiffness} matrix, $\bmath {a}=(a_0, a_1, \ldots, a_N)^{\rm T}$ is the
vector of unknowns and $\bmath {b}=(b_0, b_1, \ldots, b_{N})^{\rm T}$
is a constant vector.  The elements of $\bmath {K}$ and $\bmath {b}$
are given by
\begin{equation} 
k_{ij} = k_{ji}=\int\limits_{0}^{\frac {\pi}2} 
g_i(\theta)g_j(\theta){\rm d}\theta,  \qquad 
b_j = \int\limits_{0}^{\frac {\pi}2} 
g_j(\theta)\sigma(\theta){\rm d}\theta.  
\end{equation} 
We note that $\sigma (\theta)= \sigma (\pi- \theta)$ because of the
symmetry with respect to the equatorial plane. Since the $g_n(\theta)$
have this property as well, $n$ can take both even and odd values.
The convergence to $\sigma (\theta)$ is guaranteed if
$\lim_{N\rightarrow \infty}{\cal J}=0$.

\begin{figure*} 
\centerline{\hbox{\epsfxsize=2.1in\epsfbox{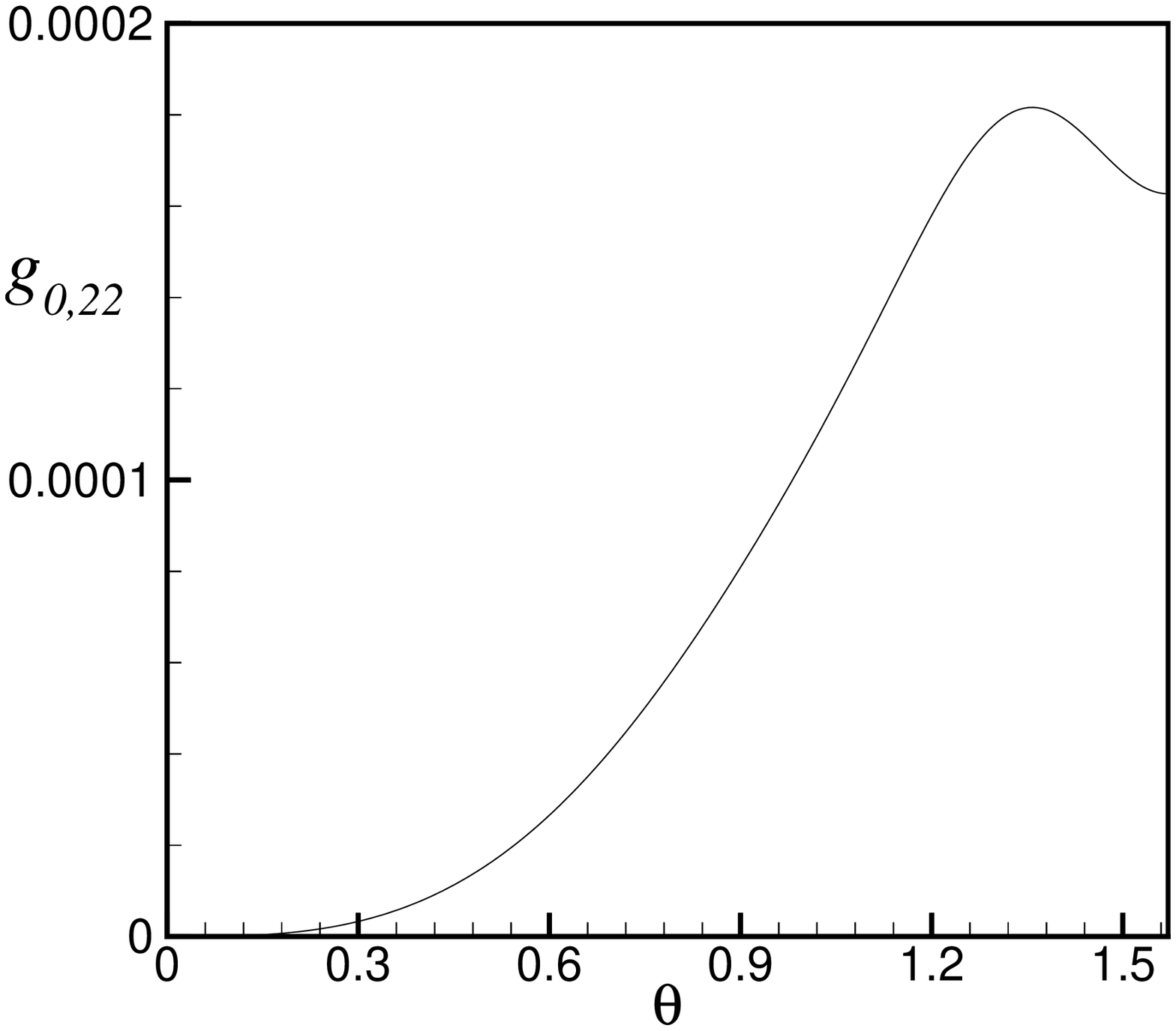}\hspace{0.3cm} 
		  \epsfxsize=2.1in\epsfbox{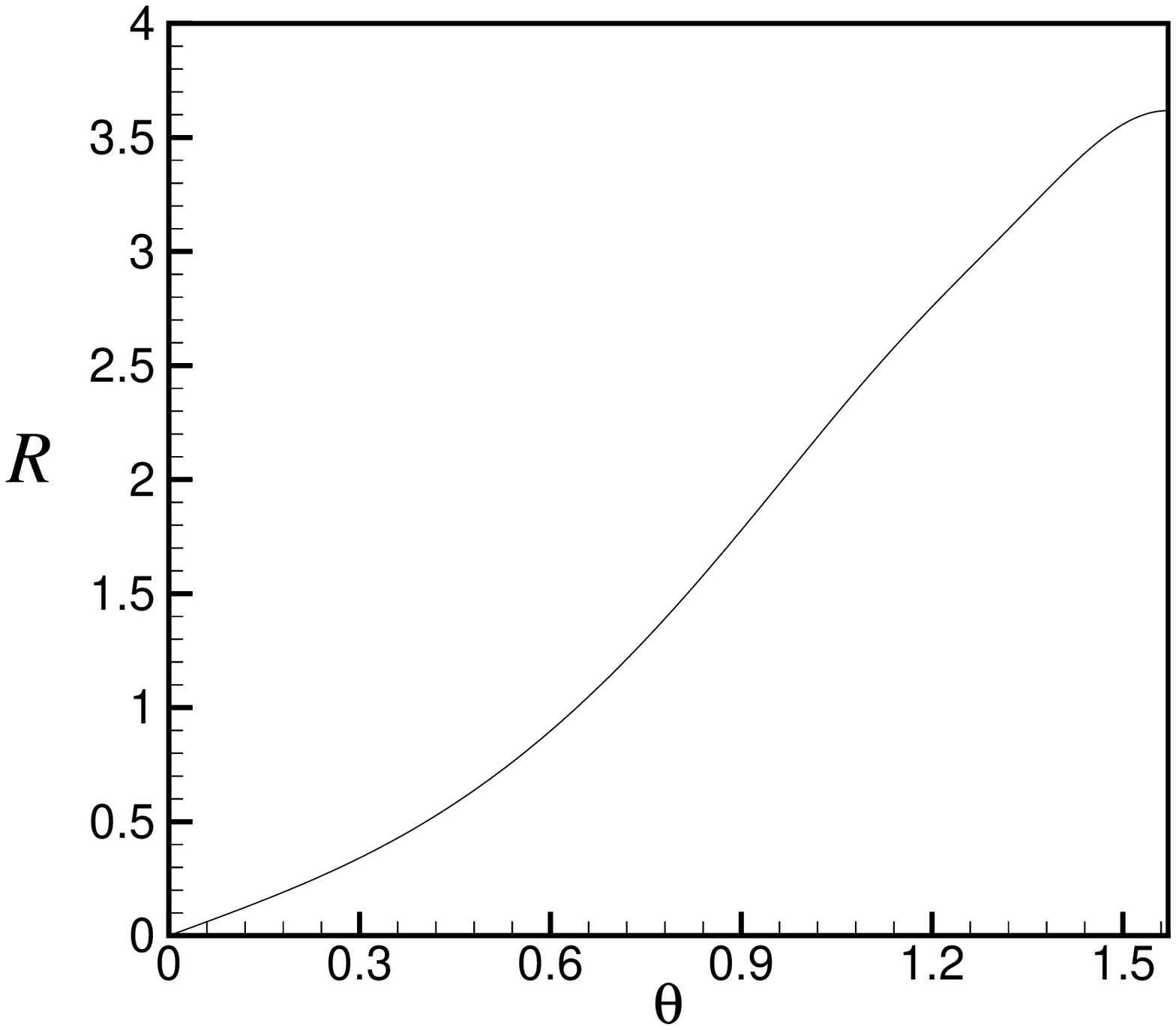}\hspace{0.3cm} 
		  \epsfxsize=2.1in\epsfbox{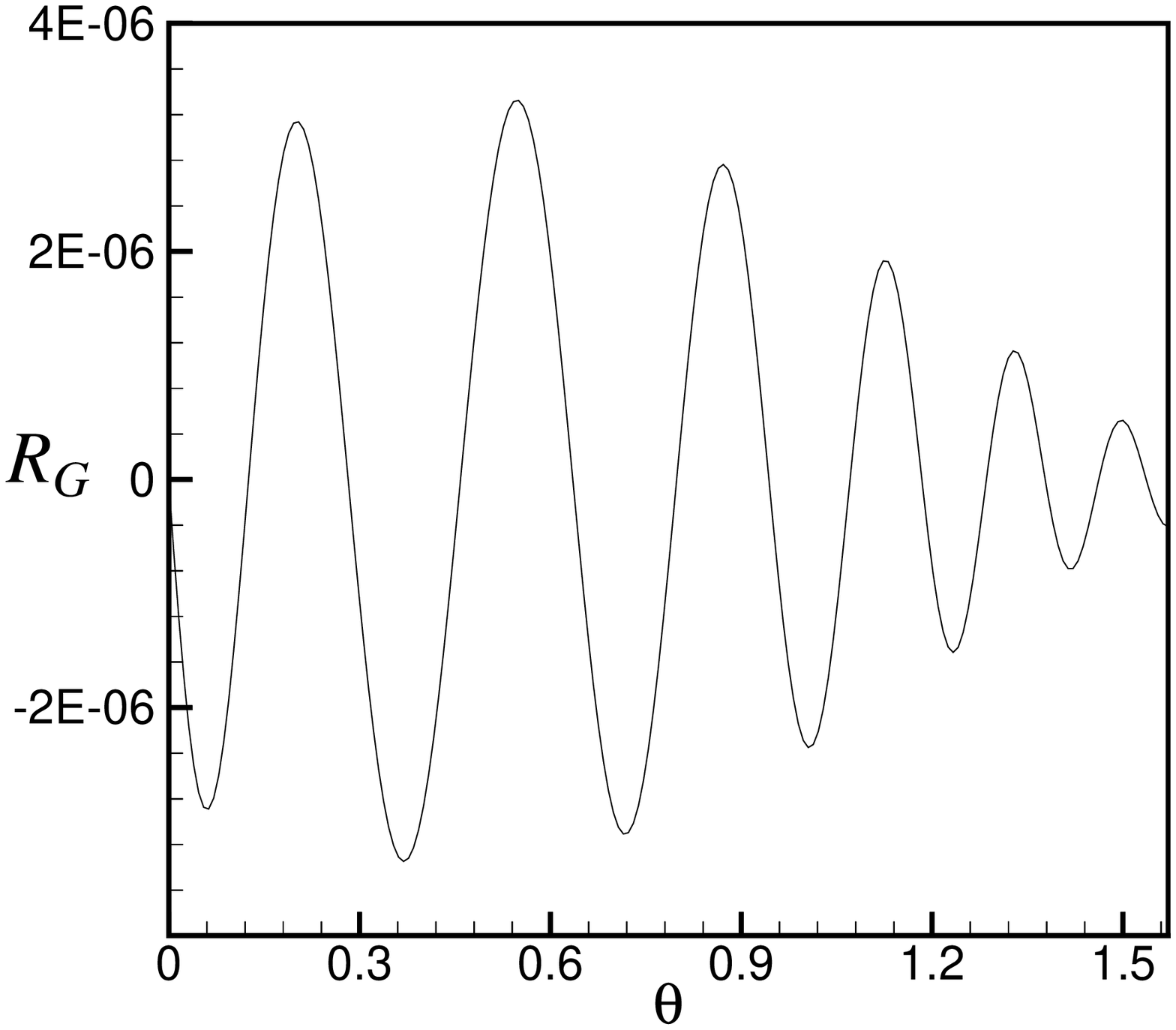}}} 
\centerline{\hspace*{3cm}$(a)$\hfill$(b)$\hfill$(c)$\hspace{1.0in}} 
\caption[Fig2]{(a) The basis function $g_{0,22}(\theta)$ for
$\gamma=0.7$ and $m=n=2$. (b) The residual function $R(\theta)$ when
$\sigma (\theta)$ is approximated by ${\cal G}_0(v,w)$. (c) The global
residual function $R_G(\theta)$.}
\end{figure*} 
 
As an example, we set $\gamma=0.7$ and follow the procedure mentioned
above. We compute $a_n$ for different values of $N$ and increase $N$
until ${\cal J}$ attains its minimum ($\approx 10^{-12}$) and the
accuracy of the numerical solutions is saturated.  Our convergence
condition is satisfied for $N=28$.  We find that
$\sigma_N(\theta)$ agrees with $\sigma (\theta)$ to an
accuracy of $10^{-6}$, which is consistent with the minimum value of
${\cal J}$. The series expansion of ${\cal G}(w)$ is also convergent
in the domain $0 \le w \le w_{\rm max}$ where we have defined
\begin{equation}
\label{eq:w-circular-orbit}
w_{\rm max}=w_+(\frac2{4-\gamma},\theta=\frac{\pi}{2}),
\end{equation}
so that $w_{\rm max}$ corresponds to the circular orbit in the
equatorial plane.  Figure 1{\em a} shows the variation of the residual
function $R(\theta)=\sigma (\theta)-\sigma_N (\theta)$ versus
$\theta$. The corresponding distribution function, which is a
well-defined positive function, has been plotted versus $\overline
w=w/w_{\rm max}$ and shown in Figure 1{\em b}.  By studying the
history of ${\cal J}$ versus $N$ (Figure 1{\em c}) it follows that the
extremum of ${\cal J}$ is indeed a minimum. Furthermore, the envelope
of the residual function is almost uniform, which indicates optimal
fitting of $\sigma (\theta)$. Our numerical experiments show that the
speed of convergence increases when $\gamma \rightarrow 0$.
  
An alternative method for the determination of the $a_n$ coefficients
would be to expand $\sigma(\theta)$ and $g_n(\theta)$ in Fourier
series, and to calculate $a_n$ by comparing the coefficients of the
sine functions on both sides of (\ref{eq:fricke-series-s}). However,
the evaluation of the Fourier coefficients considerably increases the
required computational effort, and we have not followed this
approach. Direct evaluation of $f(E, L_z)$ by means of the HQ method
is discussed in \S\ref{sec:with-bh-two}.

Our results show that, just as for the scale-free spheroids of Q95 and
the scale-free power-law galaxies of Evans (1994), the scale-free ST
models admit self-consistent $f(E, L_z)$ distribution functions.  In
the former models the function $G(w)$ is monotonic, but for the ST
models it has a maximum.

\subsection{Distribution functions $f(E, L_z, I_3)$} 
\label{sec:no-bh-three}

Dejonghe \& de Zeeuw (1988, hereafter DZ) constructed three-integral
distribution functions for Kuzmin's (1956) model using a generalized
Fricke's (1952) method. They expanded the given model density $\rho$
in terms of the potential $V$ and the polar cylindrical radius $\varpi
=\sqrt {x^2+y^2}$, and wrote the distribution function $f$ as the sum
of two parts: $f=f_2\!+\!f_3$. $f_2$ and $f_3$ are two- and three-integral
distribution functions, respectively. One can integrate $f_3$ to
obtain the corresponding density $\rho_3$, which is subtracted from
$\rho$.  The remaining density $\rho\! -\! \rho _3$ is then reproduced by
$f_2$.  DZ computed the coefficients of the Fricke expansion by direct
comparison of the series representations for $\rho$ and the integral
of $f$.\looseness=-2

We can construct three-integral distribution functions by solving
(\ref{eq:fund-int-scaled}) for ${\cal G}(v,w)$ by means of a method
developed by DZ for axisymmetric systems with St\"ackel potentials.
This can be considered as a perturbative approach in which one
exploits the existence of two-integral distribution functions. We
assume ${\cal G}$ has the form
\begin{equation} 
\label{eq:three-int-gfunction}
{\cal G}(v,w)={\cal G}_0(v,w)+{\cal G}_1(w). 
\end{equation} 
Then, we choose a specific form for ${\cal G}_0(v,w)$ and compute the
resulting density $\tilde \sigma (\theta)$. Subtracting $\tilde \sigma
(\theta)$ from $\sigma (\theta)$ leaves a residual function
$R(\theta)$. Finally, we determine the two-integral ${\cal G}_1(w)$
that is consistent with $R(\theta)$.

We first consider simple monomial forms for ${\cal G}_0$, 
\begin{equation} 
\label{eq:monomial-gfunction}
{\cal G}_0(v,w)=a_{0,mn}v^mw^n. 
\end{equation} 
Any combinations of (\ref{eq:monomial-gfunction}) can also be
chosen. By substituting (\ref{eq:monomial-gfunction}) in the
fundamental integral equation (\ref{eq:fund-int-scaled}), and carrying
out the integrations over $v$, $w$ and $u$, we find\looseness=-2
\begin{equation} 
\tilde{\sigma}(\theta)=a_{0,mn}g_{0,mn}(\theta).  
\end{equation} 
The explicit expression for $g_{0,mn}(\theta)$ is (see Appendix A)
\begin{eqnarray} 
\label{eq:g0mn-coefficients}
g_{0,mn}(\theta)\!\!\!\! &=& \!\!\!\! \sum_{i=0}^{m} \sum_{j=0}^{m-i} \!
       ~\sum_{k=0}^{m-i-j} \sum _{l=0}^{i}
      {\scriptsize \left(\!\!\!   
       \begin{array}{c} m \\ i \end{array} \!\!\!\right)} 
      {\scriptsize \left(\!\!\!  
       \begin{array}{c} m\!-\!i \\ j \end{array}\!\!\! \right)} 
      {\scriptsize \left(\!\!\! \begin{array}{c} m\!-\!i\!-\!j \\ k \end{array} 
       \!\!\! \right)}    
      {\scriptsize \left(\!\!\!\begin{array}{c} i \\l\end{array}\!\!\!\right)} 
                                                \nonumber \\		
&\phantom{=}& \!\!\!\! \times (-1)^{m-i-j-k+l} 2^{i+k+\frac {n+1}{2}} 
              {\rm B}(i+{\textstyle {1\over 2}}, {\textstyle{1\over 2}}) 
                                                \nonumber \\
&\phantom{=}& \!\!\!\! \times
     {\rm B}\left(  {\textstyle{\gamma\over 2-\gamma}}\!+\!k\!+\! 
		    {\textstyle{m+n\over 2-\gamma}},  
		     1\!+\!i\!+\!j\!+\! {\textstyle{n+1\over2}} \right)
                                                \nonumber \\
&\phantom{=}& \!\!\!\! \times \frac { {P_1(\theta)}^{m-i-2j+2k-\gamma k}
                                    {P_2(\theta)}^{n+1+2j} }
             {(1\!+\!n\!+\!2j\!+\!2l) {P(\theta)}^{\frac i{2-\gamma}}}, 
\end{eqnarray} 
where 
\begin{eqnarray} 
P_1(\theta)=\frac {1+\cos \theta}{P(\theta)^{\frac 1{2-\gamma}}}, \qquad
P_2(\theta)=\frac {\sin \theta}{P(\theta)^{\frac 1{2-\gamma}}}.  
\end{eqnarray} 
We now calculate $R(\theta)=\sigma (\theta)-\tilde \sigma (\theta)$
and solve the following equation for ${\cal G}_1(w)$
\begin{equation} 
R(\theta)=\pi \! \int\limits_{0}^{1} \! u^{2\gamma -3 \over 2-\gamma} \,
		 {\rm d}u \int\limits_{0}^{w_+(u,\theta)} \!\!\!\! 
			       {\cal G}_1(w){\rm d}w.   
\end{equation} 
The solution of this equation for ${\cal G}_1(w)$ can be inserted in
(\ref{eq:three-int-gfunction}) to determine ${\cal G}(v,w)$. The
performance of this technique depends on the initial choice of ${\cal
G}_0(v,w)$.

As an example, we construct a three-integral distribution function for
the same case discussed in \S 3.2. We take $\gamma =0.7$,
$a_{0,22}=5000$ and assume $m=n=2$. This gives us the basis function
$g_{0,22}(\theta)$ that is plotted in Figure 2{\em a}. The residual
function of this step, $R(\theta)$, is displayed in Figure 2{\em
b}. We now assume ${\cal G}_1(w)=\sum _{k=0}^{K}b_kw^k$ and find $K$
so that $R(\theta)$ is reproduced with the best available accuracy,
i.e.,
\begin{equation} 
R(\theta) \approx \tilde R(\theta)=\pi \sum _{k=0}^{K} 
b_k \int\limits_{0}^{1} \! u^{2\gamma -3\over 2-\gamma} \, {\rm d}u 
    \int\limits_{0}^{w_+(u,\theta)} \!\!\! w^k {\rm d}w.  
\end{equation} 
The constants $b_k$ are calculated by minimizing the objective
function ${\cal J}_1=\int\nolimits _{0}^{\frac {\pi}{2}}\frac 12
[R(\theta)-\tilde R(\theta)]^2 {\rm d}\theta$ with respect to the
variations of $b_k$. Numerical computation shows that ${\cal J}_1$
converges to a minimum of $5.3 \times 10^{-12}$ by taking $K=15$.  The
global residual function, $R_G(\theta)=R(\theta)-\tilde R(\theta)$, is
plotted in Figure 2{\em c}. The three-integral distribution function
obtained from (\ref{eq:three-int-gfunction}) is positive for all
possible values of $u$ and $w$.  The residual functions of the two-
and three-integral distribution functions, displayed in Figures 1{\em
a} and 2{\em c}, have similar behaviours. Their envelopes are nearly
identical, and the maximum deviation from $\sigma (\theta)$ is
approximately equal to the square root of the minimum value of the
objective function, as expected. The minimum value of the objective
function is the limit of available numerical accuracy.

\section{ST models with central black holes} 
\label{sec:with-bh}

In the presence of a central black hole, i.e., when $M\not=0$ in
equation (\ref{eq:potential-parabolic}), the ST models still have
separable potentials but lose their scale-freeness. The construction
of distribution functions by means of the series solutions of
\S\ref{sec:no-bh} becomes more complicated, even for the two-integral
case. We overcome this problem by using the contour integral method of
Hunter \& Qian (1993).

\subsection{Construction of $f(E, L_z)$}
\label{sec:with-bh-two}

The expressions for the density and potential of the ST models in
cylindrical polar coordinates ($\varpi, \phi, z$) are given in
equation (\ref{eq:dens-pot-cylindrical}). We consider the relative
potential $\Psi=-V$ corresponding to the relative energy ${\cal
E}=\Psi-\frac 12 v^2$ (Binney \& Tremaine 1987) with $v$ the modulus
of the velocity vector.

The potential $\Psi$ can be split into $\Psi=\Psi_{\rm BH}+\Psi ^*$
where $\Psi _{\rm BH}=GM/r$ and $\Psi ^*$ is the potential induced by
the density (\ref{eq:dens-pot-cylindrical}).  The value of $L_z^2$ at
a point $(\varpi, z)$ is obtained through
\begin{equation}  
L_z^2=2\varpi ^2 \left [\Psi (\varpi ^2,z^2)\! -\! {\cal E} \right ]= 
-2\varpi _c^4 \frac {{\rm d}\Psi (\varpi ^2,0)}{{\rm d}\varpi ^2} 
\vert _{\varpi =\varpi _c},  
\end{equation} 
where $\varpi_c$ is the radius of the circular orbit of energy ${\cal
E}$ in the equatorial plane, which is obtained from
\begin{equation} 
\label{eq:circular-orbit}
{\cal E}=\Psi(\varpi _c^2,0)+\varpi _c^2 \frac {{\rm d} 
\Psi (\varpi ^2,0)}{{\rm d}\varpi ^2}\vert _{\varpi =\varpi _c}. 
\end{equation} 
At a fixed energy ${\cal E}$, $L_z$ takes its maximum value for
$\varpi=\varpi _c$ and $z=0$. We denote this maximum by
$L_c=L_z(\varpi_c)$ and define $\overline w=L_z/L_c$, which is equal
to $w/w_{\rm max}$ of equation (\ref{eq:w-circular-orbit}). For the ST
models equation (\ref{eq:circular-orbit}) reads
\begin{equation} 
\label{eq:alternate-lzmax}
(4-\gamma)\varpi _c^{3-\gamma}+ 
{\cal E}\varpi _c- {\textstyle{1\over 2}} GM=0. 
\end{equation} 
It is convenient to express (\ref{eq:alternate-lzmax}) in terms of the
dimensionless parameter $\zeta=[\Psi _{\rm BH}(\varpi _c^2,0)/\vert
\Psi ^*(\varpi _c^2,0) \vert]\ge 0$. Thus, we obtain $GM=2\zeta \varpi
_c^{3-\gamma}$ and equation (\ref{eq:alternate-lzmax}) is equivalent
to 
\begin{equation}
\label{eq:radius-versus-zeta}
\varpi _c=[-{\cal E}/(4-\gamma -\zeta)]^{1\over 2-\gamma}. 
\end{equation}
It follows that $\zeta >1$ corresponds to the black hole sphere of
influence, i.e., the region where the potential of the black hole
dominates, and $0<\zeta <1$ outside this region.

\begin{figure*} 
\centerline{\hbox{\epsfxsize=5.6cm\epsfbox{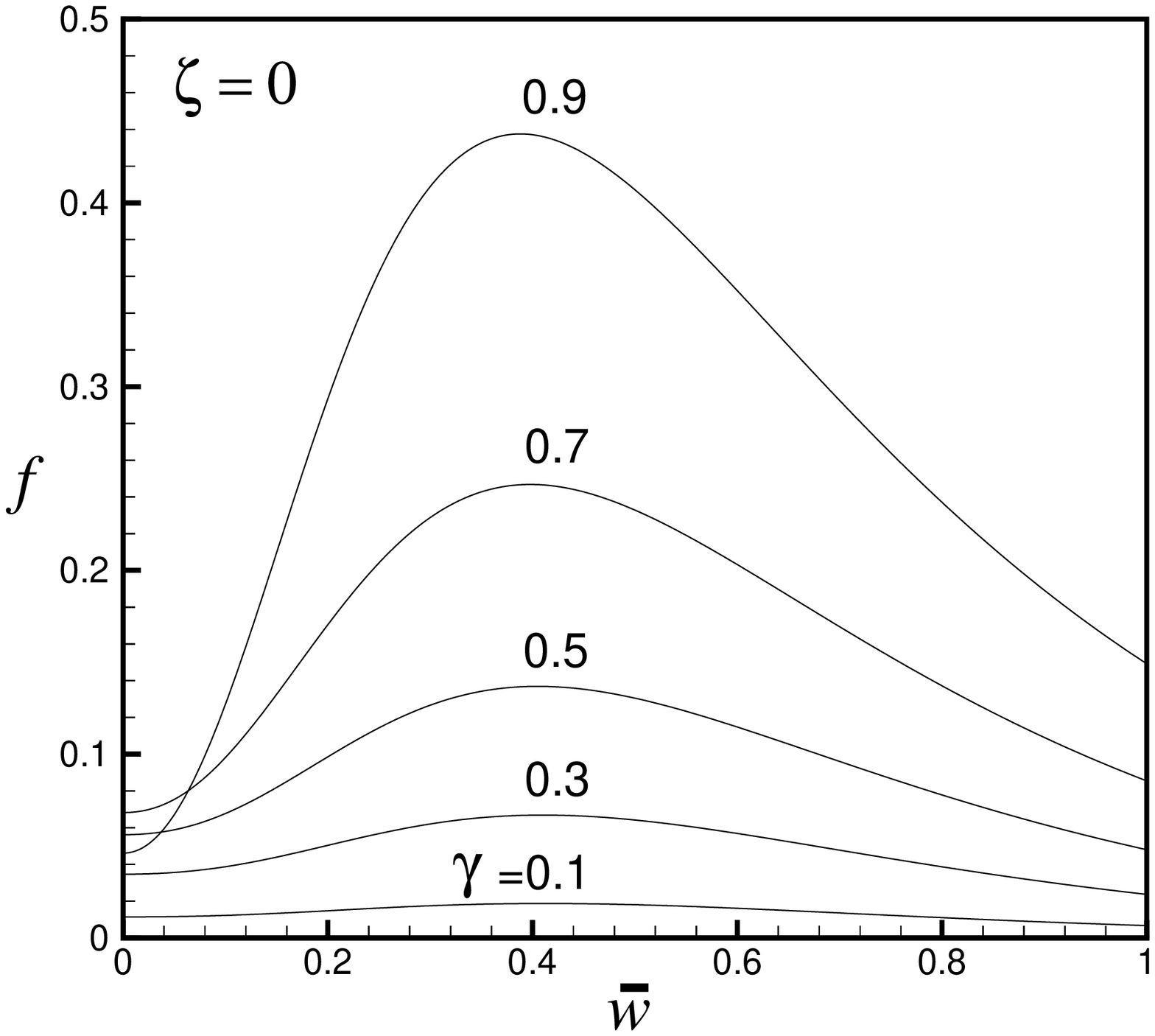}\hspace{0.3cm} 
		  \epsfxsize=5.6cm\epsfbox{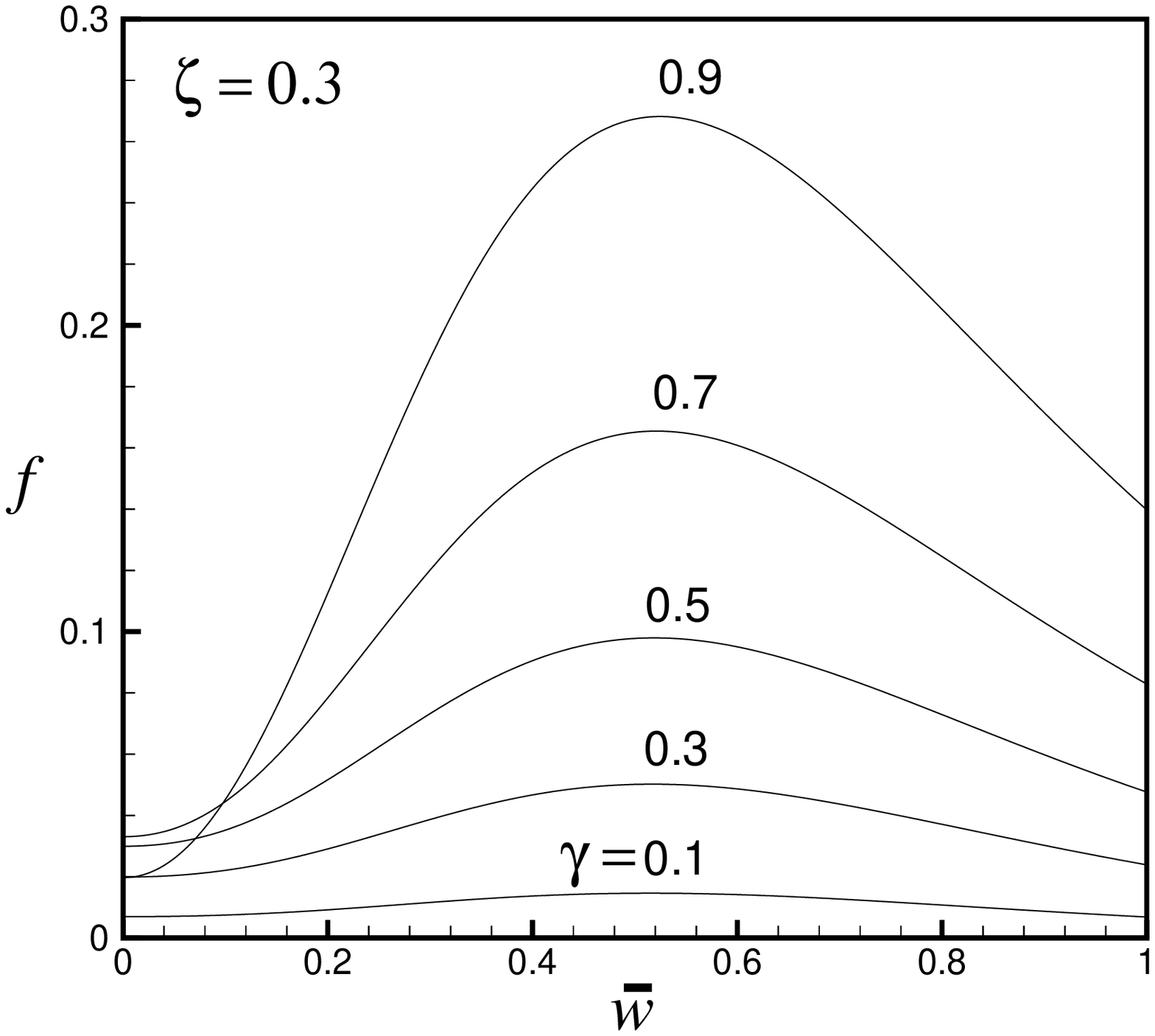}\hspace{0.3cm}
		  \epsfxsize=5.6cm\epsfbox{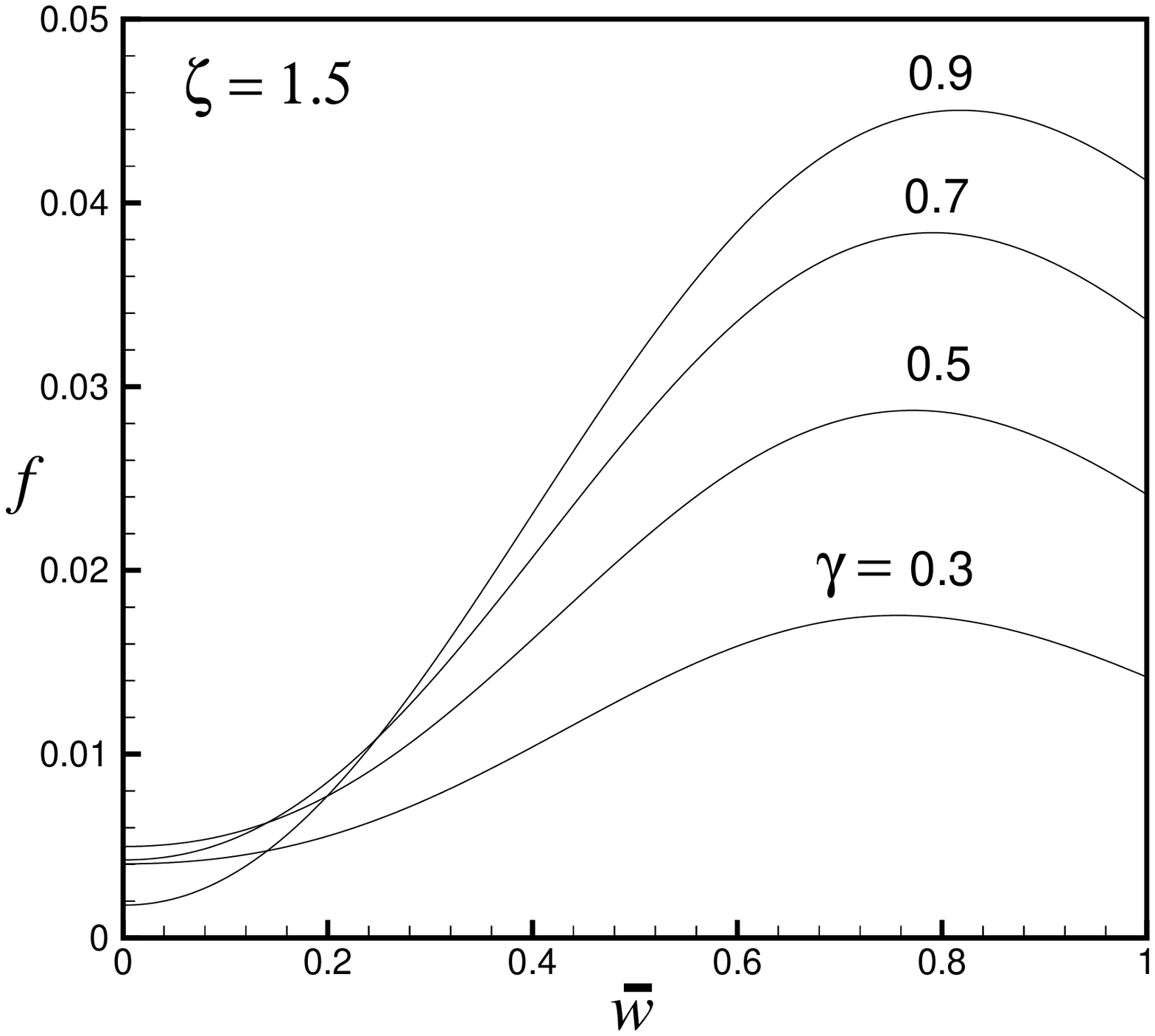}}} 
\centerline{\hspace{2.8cm}$(a)$\hfill$(b)$\hfill$(c)$\hspace{2.8cm}} 
\caption[Fig3]{Physical distribution functions constructed for ST
models using the HQ contour integral method. In all cases we have set
${\cal E}=-1$.  Panel {\em a} with $\zeta=0$ corresponds to scale-free
models in the absence of central black holes. Panel {\em b}
corresponds to $\zeta=0.3$ outside the black hole sphere of influence.
Panel {\em c} with $\zeta=1.5$ corresponds to the regions inside the
black hole sphere of influence. }
\end{figure*} 
 
To apply the HQ method, we follow Q95 and express the density $\rho$
in terms of $\Psi$ and $\varpi ^2$ as $\tilde \rho (\Psi,\varpi ^2)$
by eliminating $z^2$ between the expressions for the density and
potential given in equation (\ref{eq:dens-pot-cylindrical}).  The
contour integral solution of HQ for axisymmetric models has the form
\begin{equation} 
\label{eq:two-int-DF-hq}
f({\cal E},L_z^2)\!=\! \frac 1{4\pi ^2 {\rm i}\sqrt{2}} \!\!\!\! 
\int\limits_{\Psi _{\infty}}^{[\Psi _{\rm env}({\cal E})+]} 
  \!\!\!\!\!\!\! \tilde \rho _{11} 
    \left[\wp,\frac {L_z^2}{2(\wp\!-\!{\cal E})}\right] 
     \frac {{\rm d}\wp}{\sqrt {\wp\!-\!{\cal E}}}, 
\end{equation} 
which is evaluated in the complex $\wp$-plane. Following HQ and Q95,
we use the subscript 1 to indicate a partial derivative with respect
to the first argument of $\tilde \rho$. The symbol $[\Psi _{\rm
env}({\cal E})+]$ indicates the contour of integration, which starts
from $\Psi _{\infty}$ on the lower side of $\wp$-plane, crosses the
real axis at $\Psi _{\rm env}({\cal E})$ and ends at $\Psi _{\infty}$
on the upper side of $\wp$-plane.

For the ST models we have $\Psi _{\infty}=-\infty$ and $\Psi _{\rm
env}({\cal E})=\Psi (\varpi _c^2,0)=2\varpi _c^{2-\gamma}(\zeta
-1)$. The value of $\Psi _{\rm env}$ belongs to the interval $[\Psi
_{\rm min},\Psi _{\rm max}]$ of physically achievable potentials on
the real axis. To determine $\tilde \rho _{11}(\wp,\varpi ^2)$ we use
the implicit relation
\begin{equation} 
\label{eq:aux-relation}
\tilde \rho_{11}(\wp,\varpi ^2)=\frac {\rho _{22}(\varpi^2\!,z^2)} 
{[\Psi_2(\varpi^2\!,z^2)]^2}\!-\!\frac {\rho _2(\varpi^2\!,z^2) 
\Psi_{22}(\varpi^2\!,z^2)}{[\Psi _2(\varpi^2\!,z^2)]^3},
\end{equation} 
where each subscript 2 indicates partial differentiation with respect
to the second argument, $z^2$. For a given pair $\{\wp,\varpi
^2=L_z^2/[2(\wp-{\cal E})]\}$ on the contour of integration, the
variable $z^2$ is supplied to (\ref{eq:aux-relation}) through solving
the following nonlinear complex equation
\begin{equation} 
\label{eq:complex-relation}
{\cal P}(z^2)=\Psi \left [\frac {L_z^2}{2(\wp-{\cal E})},z^2 \right ] 
-\wp=0. 
\end{equation} 
We use the modified Newton method for solving
(\ref{eq:complex-relation}), and use the recursive formula
\begin{equation} 
\label{eq:recursion}
z^2_{n+1}=z^2_n+\frac 1{2^j}\delta, \quad  
\delta=-\left [\frac {{\cal P}(z^2)}{{\cal P}'(z^2)} \right ]_{z^2=z^2_n}, 
\quad '\equiv \frac {\rm d}{{\rm d}z^2}, 
\end{equation} 
for obtaining the $(n+1)$th estimate of the root of ${\cal P}(z^2)$.
The coefficient $1/2^j$ allows for a univariate search along the
gradient of ${\cal P}$ with the aim of taking the best step size
towards the final answer. The integer exponent $j$ is determined
through
\begin{equation} 
j={\rm min}\{k: 0\le k \le k_{\rm max}, \Vert {\cal P} 
(z_n^2+\delta/2^k)\Vert < \Vert {\cal P}(z_n^2) 
\Vert \},  
\end{equation} 
which guarantees a uniform and stable convergence. This algorithm
fails if $k>k_{\rm max}$ and we have to change the initial choice
$z_0^2$ for starting the recursion (\ref{eq:recursion}).
 
An appropriate contour is the one used by Q95, and defined by
\begin{equation} 
\wp=\Psi_{\rm env}({\cal E})+l \left [ 1\!-\!\sec \left ( \frac 
				     {\vartheta}2 \right ) \right ]          
    \!+\!{\rm i}h\sin \vartheta, \quad-\pi \le \vartheta \le \pi, 
\end{equation} 
with $l$ and $h$ positive constants. The maximum width of the
integration contour is controlled by $h$, while $l$ adjusts the
location of the local maximum of the contour.  The singularities
(poles and branch points) of $\tilde \rho_{11}(\wp,\varpi ^2)$ play an
important role in the integration along $[\Psi _{\rm env}({\cal
E})+]$.  Unfortunately, due to the implicit evaluation of $\tilde \rho
_{11}(\wp,\varpi ^2)$, we have no clear idea about the possible
singularities except for the branch point $\wp={\cal E}$ that
corresponds to $\sqrt {\wp -{\cal E}}$ in the denominator of the
integrand of (\ref{eq:two-int-DF-hq}). To gain a better sense, we
changed the width of our contour and investigated the existence of
singularities by monitoring the value of distribution function.  We
set $l=h=c\vert \Psi _{\rm env}({\cal E}) \vert$ and evaluated the
integral (\ref{eq:two-int-DF-hq}) for $c=0.2$, 0.5, 1, 5 and 10.  The
results were the same in all cases indicating that $\wp={\cal E}$ is
the only singular point on the real axis and the integrand does not
have any complex conjugate singularities.  Our computations show that
larger values of $c$ give more accurate results, and therefore we have
used $c=5$ throughout.
 
We evaluate the contour integral (\ref{eq:two-int-DF-hq}) using a
Gaussian quadrature. We carry out a change of independent variable as
$\lambda =\cos \vartheta$ ($0\le \vartheta \le \pi$) and obtain
\begin{equation} 
{\rm d}\wp=\left [ -l\frac {\sqrt {1\!-\!\lambda}} 
{\sqrt {2}(1\!+\!\lambda)} \!+\! {\rm i}h\lambda \right ] 
\frac {-{\rm d}\lambda}{\sqrt {1\!-\!\lambda ^2}}, \quad 
-1\le \lambda \le +1, 
\end{equation} 
which transforms (\ref{eq:two-int-DF-hq}) to 
\begin{equation} 
f({\cal E},L_z^2)=\frac {(a\!+\!{\rm i}b)\!-\!(a\!-\!{\rm i}b)} 
{4\pi ^2 {\rm i}\sqrt {2}}=\frac {b}{2\pi ^2\sqrt {2}}, 
\end{equation} 
where 
\begin{eqnarray}
\label{eq:abg-functions} 
a\!+\!{\rm i} b \!\!\! &=& \!\!\! \int\limits_{-1}^{+1} g(\lambda) 
\frac{{\rm d}\lambda}{\sqrt {1\!-\!\lambda ^2}}, \nonumber \\ 
g(\lambda) \!\!\! &=& \!\!\! \left[ -l\frac{\sqrt {1\!-\!\lambda}} 
					   {\sqrt {2}(1\!+\!\lambda)} 
			\!+\! {\rm i}h\lambda \right] \tilde \rho_{11} \!
\left (\wp(\lambda), \frac {L_z^2}{2[\wp (\lambda)\!-\!{\cal E}]}\right)\!. 
\end{eqnarray} 
Physical distribution functions correspond to $b \ge 0$. Assuming
$W(\lambda)=1/\sqrt {1-\lambda ^2}$ as the weight function, one can
apply the $N$-point Gauss-Chebyshev quadrature formula (Press et al.\
1992) and obtain
\begin{equation} 
\label{eq:gauss-chebyshev}
\int\limits _{-1}^{+1} g(\lambda) \frac {{\rm d}\lambda} 
{\sqrt {1-\lambda ^2}} \approx \sum _{j=1}^{N} w_j g(\lambda _j),  
\end{equation} 
where 
\begin{equation} 
w_j=\frac {\pi}N, \quad\lambda_j= \cos \left[ 
				   \frac {\pi(j\!-\!1/2)}{N} \right].  
\end{equation} 
 
We solve (\ref{eq:recursion}) with an accuracy of $10^{-10}$ and
increase $N$ until an accuracy of $10^{-8}$ is obtained in the
integration of (\ref{eq:abg-functions}) using
(\ref{eq:gauss-chebyshev}).  For instance, we constructed distribution
functions for ${\cal E}=-1$ and several values of $\zeta$ and
$\gamma$. Figure 3 shows the results for $\zeta =0$, 0.3 and
1.5. Scale-free models without central black holes correspond to
$\zeta=0$. As can be seen in Figure 3{\em a}, the graph of $\gamma
=0.7$ is in agreement with ${\cal G}(\overline w)$ of Figure 1{\em b},
which was constructed by means of the Fricke series. 

By increasing the value of $\zeta$ (moving into the black hole sphere
of influence), the maxima of $f(E,L_z)$'s are shifted to the
high-$L_z$ orbits (Figure 3). This means that sufficiently close to
the central black hole, more nearly circular orbits are needed to
maintain self-consistency in the ST models.


\subsection{Distribution functions $f(E, L_z, I_3)$} 
\label{sec:with-bh-three}
 
\begin{figure*}
\centerline{\hbox{\epsfxsize=6.5cm\epsfbox{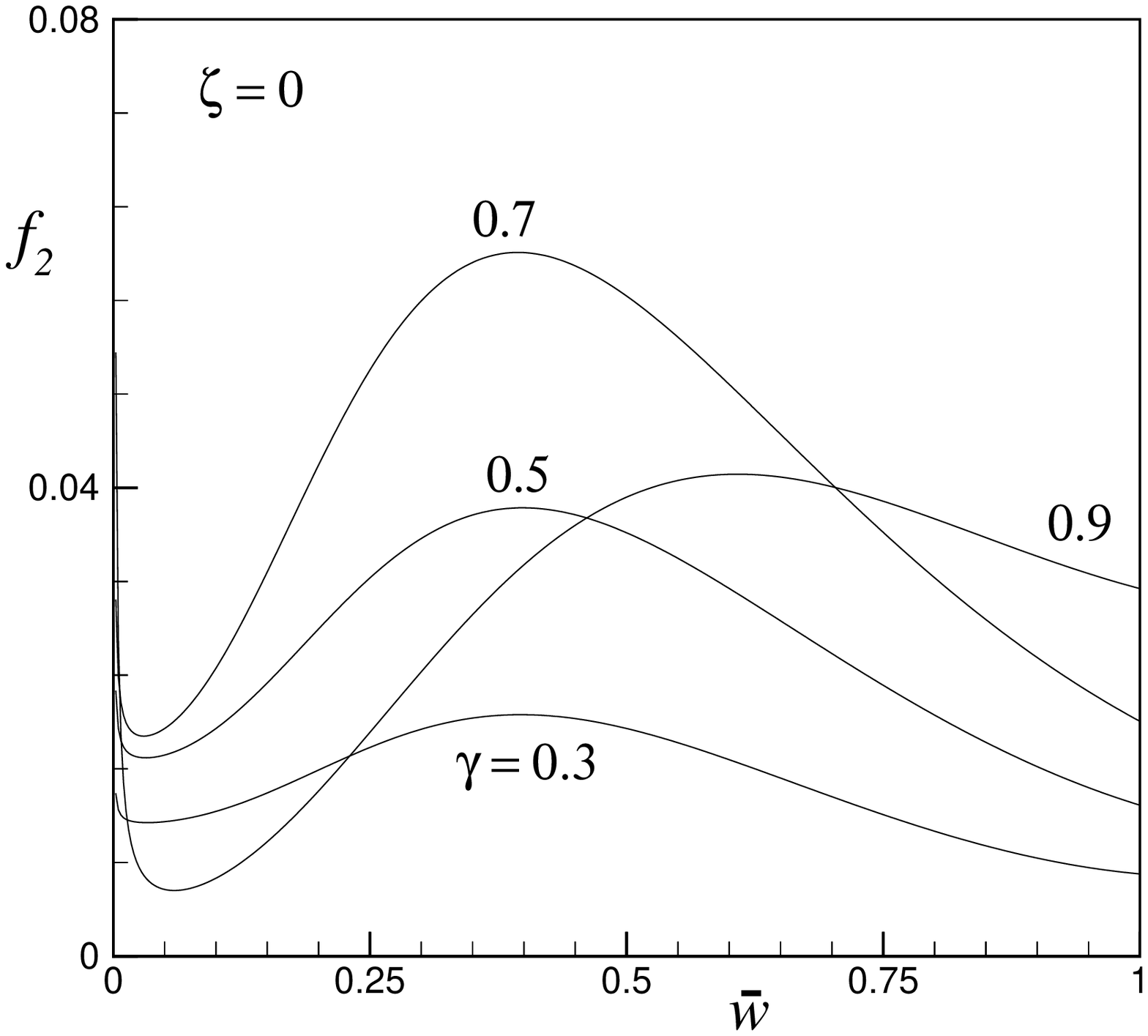}\hspace{1cm} 
		  \epsfxsize=6.5cm\epsfbox{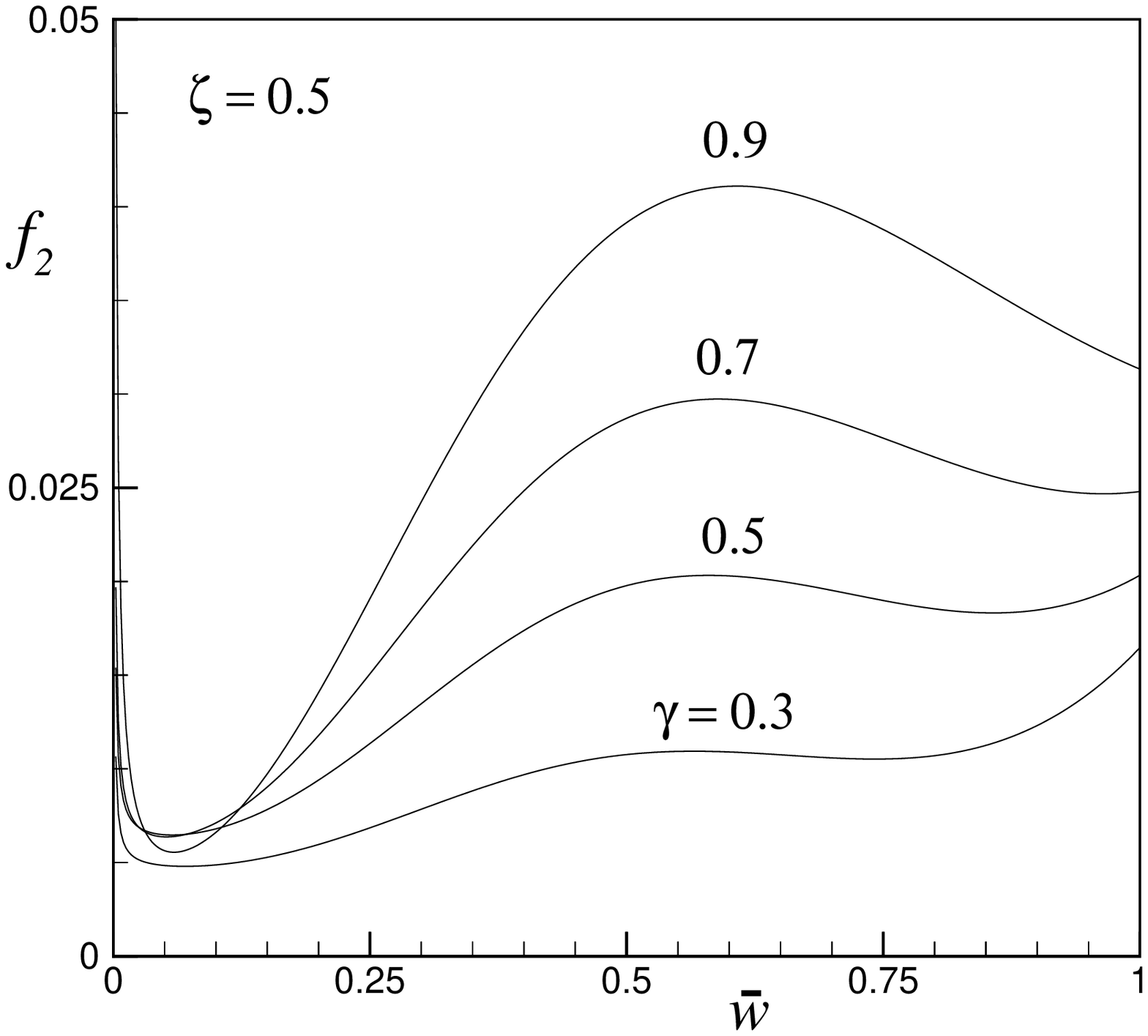}}} 
\caption[Fig4]{The two-integral $f_2$ part of
the three-integral distribution functions of ST models for
${\cal E}=-2$. The left and right panels correspond to the
scale-free ST models and ST models with central black holes,
respectively. The value of $\zeta=0.5$ is relevant to the
regions outside the black hole sphere of influence.}
\end{figure*} 

Three-integral distribution functions of ST models with central black
holes can be derived as in \S\ref{sec:no-bh} for the scale-free
models. We again write $f(E,L_z,I_3)=f_2(E,L_z)+
\epsilon f_3(E,L_z,I_3)$, assume simple forms for $f_3$ and
determine the density $\rho _3=\epsilon \int f_3 {\rm d}^3{\boldmath
v}$.  We then subtract $\rho_3$ from $\rho$ and generate $f_2$ by the
residual density $\rho _2=\rho -\rho _3$ using the method of HQ. 

For example, we assume $f_3=E^{-4}I_3^2\delta (L_z)$
where $\delta$ is the Dirac delta function. The corresponding 
density becomes (this is a consequence of eq. [\ref{eq:surf-fund-int}] 
for models with central black holes)
\begin{eqnarray}
\label{eq:three-integral-rho3}
\rho _3 \!\!\!&=&\!\!\! {\epsilon \pi \over 2 \sqrt{\xi\eta}} \Big [
{c_1^2+c_2^2 \over V^3} + {3(\xi^2+\eta^2)\over V} +
{3(\eta c_1-\xi c_2)\over V^2}       \nonumber \\
&\null& \qquad \qquad - {2 \xi \eta \over V} + {2c_1c_2\over 3V^3} +
{2(\eta c_2 - \xi c_1)\over V^2} \Big ],
\end{eqnarray} 
where 
\begin{equation}
\label{eq:rho3-parameters}
c_1 =-2 \eta ^{3-\gamma} + GM,~~
c_2 = 2\xi^{3-\gamma} - GM,
\end{equation} 
with $\xi$ and $\eta$ the parabolic coordinates. The potential
function $V$ is given by (\ref{eq:dens-pot-cylindrical}).
It is now straightforward to construct $f_2({\cal E},L_z)$ from
$\rho_2(\varpi^2,z^2)=\rho(\varpi ^2,z^2)-\rho _3(\varpi ^2,z^2)$
by following the procedure in \S\ref{sec:with-bh-two}. We have
set $\epsilon=0.02$, ${\cal E}=-2$ and have constructed
$f_2({\cal E},L_z) \ge 0$ for several choices of $\gamma$ and $\zeta$.
The results have been demonstrated in Figure 4. As this figure
shows, $f_2({\cal E},L_z)$ is oscillatory and steeply increases 
as $\overline w \rightarrow 0$. Our three-integral distribution
functions exist both in the absence and in the presence of
a central black hole for all values of $0<\gamma <1$.

\section{Discussion} 
\label{sec:conclusions}

We have constructed two- and three-integral distribution functions for
the cusped oblate models with central black holes introduced by
Sridhar \& Touma (1997a). We have computed $f(E, L_z)$ for the case
without a black hole by means of a Fricke (1952) series, and have
confirmed the result by means of the contour integral method of Hunter
\& Qian (1993). The distribution function $f(E, L_z)$ is of the form
$E^p {\cal G}(\overline w)$, with $p=-\frac{1}{2}-\frac{2}{2-\gamma}$
and $\overline w= L_z/L_c$ where $L_c(E)$ is the value of $L_z$ for
the circular orbit of energy $E$. These distribution functions are
non-negative for all values of the cusp slope $0 < \gamma < 1$, in
agreement with the results obtained by Evans (1994) for scale-free
power-law galaxies, and by Q95 for scale-free oblate spheroidal
densities. By contrast to these other models, the function ${\cal
G}(\overline w)$ is not monotonic for the strongly dimpled shape of
the ST models.

The ST models with a central black hole are not scale-free, but the HQ
contour method shows that in this case again self-consistent $f(E,
L_z)$ exist. This makes the ST models significantly different from
spheroidal cusps with central black holes, which do not admit a
self-consistent $f(E, L_z)$ when $\gamma < 1/2$ (Q95; de Bruijne et
al.\ 1996). We ascribe this difference to the dimpled shape of the ST
models, as the density distribution can be considered as the weighted
integral of two-integral components $\delta(E-E^*)\delta(L_z-L_z^*)$,
each of which have toroidal shapes. This result is unlikely to depend
on the separability of the ST models, or on the details of the orbit
structure, as the computation of $f(E, L_z)$ does not require any
knowledge of the orbits, or indeed of the existence of an exact third
integral. We speculate that the power-law galaxies of Evans (1994) can
have a physical $f(E, L_z)$ distribution function even when a central
black hole is added. As the density $\rho$ of these models is a simple
function of $\Psi$ and $\varpi^2$ (see Appendix D of Evans \& de Zeeuw
1994), these distribution functions can be found by means of the HQ
method following the procedure described in \S\ref{sec:with-bh-two}.

Nevertheless, the simple and exact form of $I_3$ in the ST models
makes it possible to construct distribution functions $f(E, L_z, I_3)$
for these models, by means of a scheme introduced by Dejonghe \& de
Zeeuw (1988). These three-integral ST models have stars on the
symmetric short-axis tube orbits as well as on pairs of reflected
banana orbits (to ensure the symmetry with respect to the equatorial
plane). The ST models have special shapes, but have a somewhat richer
dynamical structure than, e.g., the oblate models with St\"ackel
potentials in spheroidal coordinates, which contain only one major
orbit family, the short-axis tubes.

The two-dimensional versions of the ST models (Sridhar \& Touma 1997b)
are elongated discs, and no self-consistent distribution functions
exist (Syer \& Zhao 1998).  The non-self-consistency of these models
is related to the nature of the banana orbits that deposit much mass
far from the major axis where the density is maximum. Although the
meridional motions of the oblate ST models suffer this shortcoming
too, their measure in the ($E, L_z, I_3$) space is zero, and the
short-axis tubes with $L_z\not=0$ have a sufficient variety of shapes
to allow a range of self-consistent distribution functions.

Scale-free models with shallow cusps have mass distributions that
diverge strongly at large radii, and infinite projected surface
densities and stress tensors. Realistic models clearly require a
radial profile that falls off more steeply at larger radii. Q95 and de
Bruijne et al.\ (1996) have shown that at small and large radii the
scale-free models provide insight into the dynamics of these more
general models. The ST models similarly provide insight into the
nature of galactic nuclei with a central black hole and a shallow
luminosity cusp. In particular they indicate that the detailed shape
of the surfaces of constant density may have a significant influence
on the nature of the stellar velocity distribution.\looseness=-2

\smallskip
\section{acknowledgments}
MAJ thanks the Sterrewacht Leiden for hospitality, and the Netherlands
Research School for Astronomy NOVA for financial assistance.

\appendix 
\section{Determination of basis functions} 

The functions $g_{0,mn}(\theta)$ are given by  
\begin{equation} 
\label{eq:a-one}
g_{0,mn}(\theta)\!=\! \int\limits_{0}^{1} \!\!u^{2\gamma-3 \over 2-\gamma} 
	 {\rm d}u \!\!\!\! \int\limits_{0}^{w_+(u,\theta)} \!\!\!\! {\rm d}w 
		\!\!\!\! \int\limits_{v_-(u,w,\theta)}^{v_+(u,w,\theta)}  
    \frac{v^m w^n{\rm d}v} 
	 {\sqrt {[v\!-\!v_-][v_+\!-\!v]}}, 
\end{equation} 
where $v_-$ and $v_+$ are given in equation (\ref{eq:vw-int-limits}).
Define the new variable
\begin{equation} 
\mu=\frac {v-v_-}{v_+-v_-}, \qquad 0\le \mu \le 1,  
\end{equation} 
and rewrite (\ref{eq:a-one}) as 
\begin{equation} 
\label{eq:a-three}
g_{0,mn}(\theta) \!= \!\! \int\limits_{0}^{1}\!\!u^{2\gamma -3 \over 2-\gamma}
	      {\rm d}u \!\!\!\!\! \int\limits_{0}^{w_+(u,\theta)} \!\!\!\!\!\!\! 
		    w^n {\rm d}w \!\! \int\limits_{0}^{1} \! 
			\frac{[v_-\!+\!(v_+\!-\!v_-)\mu]^m {\rm d}\mu} 
			      {\sqrt{\mu (1-\mu)}}. 
\end{equation} 
The inner integral in (\ref{eq:a-three}) can be evaluated by writing
out the binomial expansion for $[v_-+(v_+-v_-)\mu]^m$, and using the
definition of the beta-function. The result is 
\begin{equation} 
\sum_{i=0}^{m} {\scriptsize \left(\!\!  
\begin{array}{c} m \\ i \end{array} \!\!\right)} 
v_-^{m-i} (v_+\!-\!v_-)^i {\rm B}(i\!+\!{\textstyle{1\over 2}}, 
				  {\textstyle{1\over2}}). 
\end{equation} 
We now collect the $w$-dependent terms of $v_-$ and $v_+\!-\!v_-$ 
and write 
\begin{equation} 
v_-\!=\!v_1(u,\theta)\!+\!w^2 \! v_2(u,\theta), \quad 
v_+\!-\!v_-\!=\!v_3(u,\theta)\!+\!w^2 \!v_4(u,\theta), 
\end{equation} 
where 
\begin{eqnarray} 
v_1(u,\theta) \!\!\!&=&\!\!\! \frac {1+\cos \theta} 
{P(\theta)^\frac {3-\gamma}{2-\gamma}} 
[2u(1\!+\!\cos \theta)^{2-\gamma}\!-\!P(\theta)] u^{\frac {1}{2-\gamma}},
						     \nonumber \\  
v_2(u,\theta) \!\!\! &=& \!\!\! u^{-\frac {1}{2-\gamma}} 
	 \frac{P(\theta)^{\frac 1{2-\gamma}}}{2(1\!+\!\cos\theta)}, \nonumber \\ 
v_3(u,\theta) \!\!\! &=& \!\!\! 2u^{\frac 1{2-\gamma}}(1\!-\!u) 
			  P(\theta)^{-\frac 1{2-\gamma}}, \nonumber \\
v_4(u,\theta) \!\!\! &=& \!\!\! -u^{-\frac 1{2-\gamma}} 
	    \frac {P(\theta)^{\frac 1{2-\gamma}}} {\sin ^2 \theta}. 
\end{eqnarray} 
By combining these results, and carrying out the integration over $w$, we
obtain
\begin{eqnarray} 
g_{0,mn}(\theta) \!\!\! &=& \!\!\!\! \int\limits_{0}^{1} 
			\!\! u^{2\gamma -3\over 2-\gamma} {\rm d}u 
    \sum _{i=0}^{m} \sum_{j=0}^{m-i} \sum _{l=0}^{i}
   {\scriptsize \left(\!\! 
	   \begin{array}{c} m \\ i \end{array} \!\!\right)}  
   {\scriptsize \left(\!\! 
           \begin{array}{c} m\!-\!i \\ j \end{array} \!\!\right)}
   {\scriptsize \left(\!\! 
           \begin{array}{c} i \\ l \end{array} \!\!\right)}    \nonumber \\
&\phantom{=}& \!\!\!\! \times 
   {\rm B}(i+{\textstyle{1\over2}},{\textstyle{1\over 2}}) 
    v_1^{m-i-j} v_2^j v_3^{i-l} v_4^l                          \nonumber \\
&\phantom{=}& \!\!\! \times
    \frac {w_+(u,\theta)^{n+1+2j+2l}}{n+1+2j+2l}. 
\end{eqnarray} 
Finally, we replace $v_1^{m-i-j}$ with its binomial expansion, collect
the $u$-dependent terms and then integrate over $u$. This gives us
equation (\ref{eq:g0mn-coefficients}).
 
\end{document}